\documentclass[reprint,amsmath,amssymb,aps,pre]{revtex4-2}
\usepackage[T1]{fontenc}
\usepackage[utf8]{inputenc}
\usepackage{mathptmx}
\usepackage[scaled=0.85]{beramono}

\usepackage{tabularx}  
\usepackage{booktabs}
\usepackage{amsmath, amssymb, amsfonts, amsthm, bm, mathtools, siunitx}

\usepackage{graphicx}
\usepackage{dcolumn, booktabs}
\usepackage{multirow}

\usepackage[dvipsnames]{xcolor}
\usepackage{soul} 

\usepackage[colorlinks=false, hidelinks]{hyperref}

\usepackage{url}

\usepackage[version=3]{mhchem}
\usepackage{chemformula}

\usepackage[toc,page]{appendix}
\usepackage{natbib}

\usepackage{amsfonts}

\newcommand*{\blauw}[1]{{{#1}}}

\begin{document}

\hbadness=99999
\vbadness=99999
\hfuzz=0.5pt
\vfuzz=5pt

\preprint{APS/123-QED}

\title{Data-driven reconstruction of dynamical systems using Takens' Theorem, manifold learning, and universal function approximators}

\author{Maximilian Topel}
\affiliation{Department of Physics, University of Chicago, Chicago, IL 60637, United States of America}
\affiliation{Department of Applied Mathematics, Northwestern University, Evanston, IL 60208, United States of America}

\author{Andrew L. Ferguson}
\email{andrewferguson@uchicago.edu}
\affiliation{Pritzker School of Molecular Engineering, University of Chicago, Chicago, IL 60637, United States of America}
\affiliation{Department of Chemistry, University of Chicago, Chicago, IL 60637, United States of America}

\begin{abstract}
\noindent Embedding theorems can be used to provide theoretical guarantees about the relation between low-dimensional observations of a system and its full-dimensional state and dynamics. Such theorems do not, however, provide guidance on observable choice, embedding construction, or methodologies to learn the mapping between the embedding and full-dimensional state. In this work, we develop an algorithmic framework, TAkens Reconstruction (TAR), to analyze and reconstruct arbitrary dynamical systems from low-dimensional time series using an integration of Takens' Delay Embedding Theorem, manifold learning techniques, and universal function approximators. We validate TAR in applications to a variety of simulated and observed dynamical systems and use it to investigate how delay vector structure impacts reconstruction accuracy. In an ecological system, we show that simple predator-prey dynamics can be reconstructed with observations taken over a wide variety of embedding time scales. In molecular dynamics simulations of the protein Villin, we demonstrate how including multiple time delays of the same observable series can be used to improve reconstruction of systems with multiple characteristic time scales. In the trade record of Vanguard S\&P 500, we show how the approach exposes underlying dynamical phenomenologies in the data and accurate return predictions over short time horizons without access to full-dimensional market observations. We develop and release an open-source software package to enable the application of TAR to arbitrary dynamical systems. 
\end{abstract}


\maketitle

\section{\label{TARIII:sec:Intro}Introduction}

Dynamical systems, be they collections of molecules, colonies of bacteria or groups of animals, are often modeled via a reductionist perspective by propagating the evolution of each constituent element of the system under its interactions with all of the others. The state of the system can be specified by the phase space coordinates serving as inputs to these equations and the accessible phase space is defined as those regions of coordinate space that the dynamical system can visit under sufficiently long trajectories. The full-dimensional representation of a given dynamical system is defined by a vector of the phase space coordinates that are inputs to the deterministic or stochastic propagation equations. 
In the case of computer simulations of a molecular system, for example, we can completely describe the dynamical system state by recording the positions and velocities of all constituent atoms as we propagate them in time under Hamiltonian mechanics using molecular dynamics (MD) algorithms \cite{Hollingsworth2018}. In experiments observing molecules evolving over time, however, access to all atomistic positions and velocities is typically not available \cite{Wu1994,Roy2008rr}. The problem becomes even more pronounced when considering biological systems, where physics-based calculations are intractable, or to social systems, where the underlying rules are unknown. We can, however, collect time series observations of most dynamical systems. A central objective of dynamical systems theory is to learn a mapping between low-dimensional observations and the full-dimensional representation of the dynamical state of the system \cite{Sugihara2011PLOS,Solijacic2022,Suddhasattwa2023,keverkidis2017}.

Limits on the nature of the relationship between time series and full-dimensional dynamics have been thoroughly explored analytically \cite{Takens,Cao1998,stark1999delay,Broomhead1986,stark2003delay,probtak}. Takens' Delay Embedding Theorem states that, under mild technical conditions, there exists a bijective map between vectors constructed from time series observables and the full-dimensional representation of a system \cite{Takens}. Stark and Broomhead elaborated upon this work to demonstrate that this map exists even for systems under stochastic or deterministic forcing \cite{stark1999delay,stark2003delay,Broomhead1986}. Cao \textit{et al.}\ furthered Takens' univariate work by extending the mathematical principle to integrate information from multiple observables \cite{Cao1998}. Bara\'nski \textit{et al.}\ showed that Takens' Theorem can be extended to a non-dynamical probabilistic embedding theorem \cite{probtak}.  While theoretical guarantees exist regarding the relation between low-dimensional observations and full-dimensional system representations, Takens' Theorem provides no commentary on the form of this map. Moreover, the theorem pertains to systems for which we have a complete sampling of a phase space, giving little guidance on its utility in empirical studies where it is unknown if accessible phase space is fully sampled. The accessible observables and time series may not be arbitrarily definable, but rather imposed by practical limitations. These considerations motivate empirical study of optimal data structuring to best represent system dynamics and reconstruct the dynamical system state.

With the advent of artificial neural networks as practical universal functional approximators \cite{Hassoun1996}, it has become possible to readily learn these \textit{a priori} unknown mappings in systems where these relations are not available analytically. In prior work, we have combined the mathematical guarantees on diffeomorphic mappings between vectors constructed from arbitrary low-dimensional observations and full-dimensional representations provided by Takens' Theorem with data-driven learning of empirical universal functional approximators capable of estimating their forms in applications to molecular systems \cite{Ferg16,Ferg18,topel2020,topel2023,topel2026}. The central question we seek to address -- how to optimally reconstruct full-dimensional dynamics from low-dimensional observations for arbitrary dynamical systems -- can be divided in two parts. First, there is a methodological question regarding how one can perform these reconstructions. Second, there is a practical question of how one should choose and structure the time series of these observables used as inputs to reconstruction pipelines. In this work, we engage both of these challenges in the reconstruction of dynamical system states from low-dimensional observables in simulated and observed dynamical systems. In numerical simulations of a Lotka-Volterra predator-prey model \cite{LV}, we demonstrate the effect of delay choice in reconstruction quality.  In molecular dynamics simulations of the protein Villin, we show how structuring time series observables over multiple delay times can improve the reconstruction of systems with multiple characteristic time scales. In the trade record of Vanguard S\&P 500, we demonstrate learning of the structure of the phase space of price changes from historical returns time series. We also develop an accompanying software package, TAkens Reconstruction (TAR), to perform learning and reconstruction of arbitrary dynamical systems from time series data that we make available for free and open source download from \url{https://github.com/Ferg-Lab/TAR} and via the persistent Zenodo DOI \href{https://doi.org/10.5281/zenodo.20115543}{10.5281/zenodo.20115543}.


\section{\label{TARIII:sec:Dynamics}Reconstructing Dynamical System States from Low-Dimensional Time Series}

The propagation of a dynamical system may be represented numerically by a set of coupled differential equations or, equivalently, a matrix operating upon a state space vector that defines the system's position in phase space \cite{Brin_Stuck_2002,strogatz}. The state space of an $N$ dimensional dynamical system can be written as $\mathcal{S} \subseteq \mathbb{R}^{N}$ with state $\mathbf{r}(t)\in\mathcal{S}$ at time, $t$. Typically, individual degrees of freedom of a dynamical system will interact with one another and these couplings lead $\mathcal{S}$ to admit a lower dimensional representation, $\mathcal{M} \subseteq \mathbb{R}^{k}$, where $k$ is the intrinsic dimensionality of the system \cite{RuelleEckmann,sauer1991embedology,ISOMAP,LLE,coifman2006diffusion,Coifman24052005}. This intrinsic dimensionality of the system, the minimum dimensionality required to accurately describe the dynamics of our system, is generally much smaller than $N$ \cite{RuelleEckmann,sauer1991embedology}.  For example, the evolution of a molecular system composed of $n$ atoms, $\mathcal{S}\subseteq \mathbb{R}^{N}$, proceeds in a $N=6n$ dimensional space comprising the positions and velocities of the $n$ constituent atoms. However, it has long been understood that the long time dynamical evolution can be adequately described in a low dimensional space of leading collective variables where $\mathcal{S}$ admits a low dimensional representation $\mathcal{M} \subseteq \mathbb{R}^{k}$ where $k\ll6n$ \cite{garcia1992large,amadei1993essential,hegger2007complex,zhuravlev2009deconstructing,das2006low,ferguson2010systematic,topel2023}.

\subsection{\label{TARIII:sec:Embeddings}Takens' Delay Embedding Theorem}

The (weak) Whitney Embedding Theorem guarantees that any continuous function $\mathcal{F}(t)$ evolving on an $N$ dimensional manifold $\mathcal{S} \subseteq \mathbb{R}^N$ can be approximated by a smooth embedding into $(2N+1)$ dimensions \cite{whitney1936differentiable,difftop}. However, for dynamical systems where some low dimensional attractor exists, we can extend this embedding principle to restrict the dynamics on $\mathcal{S}$ to some low dimensional phase space $\mathcal{M} \subseteq \mathbb{R}^k$ where $k\ll N$. Interactions between individual components of $\mathcal{S}$ mean that time series observations of each degree of freedom in a dynamical system may contain information about other degrees of freedom and, consequently, the intrinsic dimensionality of the dynamics on $\mathcal{S}$ can be described in reduced dimension $\mathbb{R}^{k}$ \cite{RuelleEckmann,sauer1991embedology}. As such, there exists an ``intrinsic manifold'' $\mathcal{M}\subseteq\mathbb{R}^{k}$ that then embeds the dynamics on $\mathcal{S}$ in a latent space of dimensionality $k$ bounded by $k$ $\leq$ $d_S$, the box counting dimension of $\mathcal{S}$ \cite{difftop}. This intrinsic manifold contains the long-time dynamical evolution of the system and can be represented by a $(2k+1)$-dimensional embedding.

High dimensional dynamics can, therefore, generally be embedded in reduced dimensional space, but how can these dynamics evolving in $\mathbb{R}^k$ with $k\geq1$ be recovered from univariate time series? Time series offer a window into both the evolution of a system in phase space and the propagators driving these systems. Although a univariate time series $\mathcal{O}(t) \in \mathbb{R}^1$ evolves in a space of equal or lesser dimensionality to the intrinsic system dimensionality, the trajectories that system takes through $\mathbb{R}^1$ are generated by the underlying dynamical propagator of the system and they may be analyzed to provide information on the full-dimensional system state. 

Takens' Delay Embedding Theorem, which holds for both autonomous dynamical systems and those subject to stochastic and deterministic forcing \cite{Broomhead1986,stark1999delay}, can be viewed as an elaboration on the Whitney Embedding Theorem for time series of a single observable \cite{Takens}. The theorem states that, if properly arranged, scalar time series data observed over a sufficiently densely sampled and sufficiently long system trajectory are diffeomorphic (i.e., related by a smooth and invertible transformation) to the dynamics in the full-dimensional phase space \cite{Takens}. More precisely, any generic observable, $\mathcal{O}(t)$, of a dynamical system that does not contain any spurious symmetries not present in the full-dimensional system can be used to construct a time delay vector $y(t) = [\mathcal{O}(t),\mathcal{O}(t-\tau),\ldots,\mathcal{O}(t-2m\tau)] \in \mathbb{R}^{2m+1}$ with delay time $\tau$ and intrinsic dimensionality $m$ which enjoy a unique relation with the full-dimensional representation of that system's state space $\mathbf{r}(t)\in \mathcal{S}\subseteq\mathbb{R}^{N}$. This relation has three key components that are exploited in this work:
(i) the dynamics of $y(t)$ and $\mathbf{r}(t)$ are  $C^{1}$-equivalent up to observable symmetry (i.e. identical up to a smooth and continuous mapping up to observable invariance or degeneracy),
(ii) $y(t)$ must uniquely specify the state of the system (i.e. $y(t)$ is defined by at least $k$ independent variables that unambiguously define system configuration over all $t$), and
(iii) $y(t)$ lies on manifold $\mathcal{M}^{\prime}$ which is related by a smooth, invertible and bijective mapping (i.e. diffeomorphism) to the manifold $\mathcal{M}$ that we denote $\Theta:\mathcal{M}^{\prime}\rightarrow\mathcal{M}$.

There are also some important and well-known caveats \cite{Takens,stark1999delay,stark2003delay,sauer1991embedology,Cao1998,packard1980,kantz2004nonlinear,martin2024robust,topel2020}. Given the $C^{1}$-equivalence of the dynamics of $y(t)$ and $\mathbf{r}(t)$ and the diffeomorphic relation between $\mathcal{M}^{\prime}$ and $\mathcal{M}$, these manifolds are topologically identical, but are not guaranteed to be topographically identical. The diffeomorphism induces squashing and/or stretching of $\mathcal{M}^{\prime}$ that is guaranteed to preserve continuity and connectivity (i.e., the diffeomorphism may not rip the manifold or stitch it back together in new ways) and integrated probabilities, but the probability distributions themselves may be deformed by the mapping such that barrier heights and well depths are altered. 
To ensure that $y(t)$ uniquely specifies the state of the system, we must collect sufficient observations such that we well sample the phase space accessible to our dynamical system. Observables possessing some type of symmetry (e.g. translation, rotation, chiral, permutation) will not allow for the construction of delay vectors able to resolve the difference between symmetric states. The delay time $\tau$ may not be an integer multiple of a period of the dynamical evolution so as not to introduce temporal aliasing. 

Taken together, Takens' Theorem presents the theoretical foundations under which one may seek to reconstruct the dynamical state of a system from historical time series data \cite{Takens} and may be profitably integrated with data-driven numerical tools to empirically learn and approximate this mapping.

\subsection{\label{TARIII:sec:dimred}Dimensionality Reduction and Manifold Learning}

Manifold learning techniques can be used to learn a map from high-dimensional representations of dynamics in $\mathcal{S} \subseteq \mathbb{R}^N$, to the low-dimensional intrinsic manifold $\mathcal{M} \subseteq \mathbb{R}^{k}, k \ll N$ containing the leading long-time dynamics \cite{coifman2006diffusion,Coifman24052005,LLE,ISOMAP,NLPCA}. Many of these techniques rely on a particular distance metric defined between points in the high dimensional space to construct a lower-dimensional vector space containing the intrinsic manifold \cite{coifman2006diffusion,Coifman24052005,ISOMAP,LLE,NLPCA}. The choice of distance metric and of manifold learning algorithm can have a significant effect on the shape and structure of the learned latent spaces. We can recover the intrinsic system dimensionality of time series data via, for example, Cao's E1(d) method \cite{Cao97} and construct low-dimensional embeddings using manifold learning techniques such as diffusion maps, Isomap, Locally Linear Embedding (LLE), or nonlinear principal components analysis (NLPCA) \cite{coifman2006diffusion,Coifman24052005,ISOMAP,LLE,NLPCA}. Combining this manifold reduction technique with embedding theorems allows us to harness the information gains from ordering time series data via Takens' Theorem while producing the lowest dimensional representation that embeds the full-dimensional dynamics using manifold learning.

In this work, we apply the diffusion map nonlinear manifold learning technique to perform dimensionality reduction of full-dimensional training trajectories and time delayed vectors constructed from specific observables on to learned low-dimensional manifolds \cite{coifman2006diffusion,Coifman24052005,nadler2006advances,ferguson2011cpl}. Diffusion maps perform a spectral decomposition of a diffusion operator on a high dimensional data set to produce an embedding into a lower dimensional space in which, under some mild assumptions on the system dynamics, Euclidean distances in the low-dimensional embedding approximate diffusion distances in the high-dimensional space \cite{Coifman24052005,nadler2006advances}. This favorable property of the latent space embeddings means that states that have high transition probabilities are embedded close to one other and those that have low transition probabilities are embedded further apart. As such, the latent space tends to preserve the topology and connectivity of the high dimensional phase space. Furthermore, diffusion maps also provide a straightforward means to estimate the intrinsic dimensionality through a spectral gap \cite{nadler2006advances,lpbeltrami} and are robust to noise \cite{coifman2006diffusion}.

\section{\label{TARIII:sec:delayvecs}Construction of Delay Vectors}

Given full-dimensional observations of a dynamical system, manifold learning can be employed to infer the intrinsic dimensionality and approximate the intrinsic manifold. For many experimental dynamical systems, it may not be possible to observe the full high-dimensional dynamics of the system, but it may be possible to record a time series, $\mathcal{O}^j_t \in \mathbb{R}^{1}$, in one or more system observables $j$ recorded at times $t$. Such observations may measure quantities related to individual elements of a system such as the price of a stock, multiple elements of a system such as the distance between two atoms in that molecule, or measures of continuum dynamics such as the velocity of a fluid flow. It is the remarkable assertion of Takens' Theorem that the full-dimensional dynamical state of the a system may be reconstructed from delay embeddings of low-dimensional time series in one or more observables \cite{Takens}.

\textbf{Univariate time series:} We can supplement a single time series observation $\mathcal{O}^j_t$  with observations from previous times to collect histories of these observables into delay vectors as originally proposed by Takens \cite{Takens}, 
\begin{equation}
\mathcal{Y}^{j}_t = \left [ \mathcal{O}^j_t,\mathcal{O}^j_{t-\tau},...,\mathcal{O}^j_{t-(p-1)\tau} \right],
\end{equation}
where $\tau$ is the time delay between observations and $p$ is the dimensionality of the delay vector. Conceptually, we can think of a system history as a specification of the path our system takes through phase space, and delay vectors as trajectory snippets or ``micro-trajectories,'' which, under some mild conditions, uniquely define where on the high dimensional manifold $\mathcal{S}\subseteq\mathbb{R}^{N}$ the observation lies at time $t$ and discriminating between identical instantaneous values of $\mathcal{O}^j_t$. 

Mathematically, it is the remarkable assertion of Takens' Theorem that a delay embedding $\mathcal{Y}^{j}_t = \left [ \mathcal{O}^j_t,\mathcal{O}^j_{t-\tau},...,\mathcal{O}^j_{t-(p-1)\tau} \right]$ with delay time $\tau$, delay embedding dimensionality $p$, where $p$ is more than twice the intrinsic dimensionality of the system, in a generic observable $\mathcal{O}^j$ that does not contain any spurious symmetries not present in the system itself uniquely specifies the instantaneous state of the system, endows $\mathcal{Y}^{j}_t$ with a dynamical evolution is $C^1$-equivalent to that of the dynamical evolution of the system $\mathbf{r}(t)$ in the full-dimensional ambient space of all degrees of freedom, and defines these dynamics to proceed on a manifold $\mathcal{M}^{\prime}$ that is a topologically-identical image of $\mathcal{M}$ related by a smooth, invertible, and bijective (i.e., diffeomorphic) mapping \cite{Takens,stark1999delay,stark2003delay,sauer1991embedology,Cao1998,packard1980}. 

In general, the functional form of the diffeomorphism is not known, meaning that the transformation to $\mathcal{M}$ is not directly accessible, but the theorem guarantees that $\mathcal{M}$ can be constructed from $\mathcal{M}^{\prime}$ by stretching and squashing but that tearing and stitching back together are forbidden. This topological equivalence assures that distributions and pathways over $\mathcal{M}^{\prime}$ are smooth, continuous images of those over $\mathcal{M}$ and presents theoretical guarantees for empirical approximation of the diffeomorphism by data-driven learning \cite{Ferg16,Ferg18,topel2020,topel2023}. In prior work, we have demonstrated that manifold learning may be applied to full-dimensional observations of $\mathbf{r}(t)$ to approximately recover $\mathcal{M}$ and to delay embeddings $\mathcal{Y}^{j}_t$ to approximately recover $\mathcal{M}^\prime$, and regression techniques used to approximate the diffeomorphic function linking $\mathcal{M}$ and $\mathcal{M}^\prime$ \cite{topel2020,topel2023}.

\textbf{Multivariate time series:} In many experiments, it is possible to multiplex observations and record multiple observations simultaneously.  Given multiple time series, we can construct a multivariate time delay vector as proposed by Cao \textit{et al.}\ \cite{Cao1998}. For example, in the case of two data streams, we may construct $\mathcal{Y}^{i,j}_t$ by concatenating $\mathcal{Y}^{i}_t$ and $\mathcal{Y}^{j}_t$,
\begin{equation}
     \mathcal{Y}^{i,j}_t = [\mathcal{O}^i_t,\mathcal{O}^i_{t-\tau^i},..., \mathcal{O}^i_{t-(p^i-1)\tau^i}, \mathcal{O}^j_t,\mathcal{O}^j_{t-\tau^j},...,\mathcal{O}^j_{t-(p^j-1)\tau^j } ],
\end{equation}
where $\tau^i$ and $\tau^j$ and $p^i$ and $p^j$ are, respectively, time delays and delay embedding dimensionalities for observable $\mathcal{O}^{i}$ and $\mathcal{O}^{j}$.  Multivariate data streams allow for information from various sources with potentially different symmetries, relaxation times, noise susceptibilities, and responsiveness to particular aspects of the system evolution to be combined to provide a more robust and accurate estimation of $\mathcal{M}^{\prime}$ by manifold learning.

\textbf{Multi-temporal dynamics:}  Many real world dynamical systems may evolve on multiple time scales. Separability of dynamics occurring on varying time scales implies that we cannot learn information about separate dynamics occurring at time scales $\mathcal{T_{\alpha}}$ $\ll$ $\mathcal{T_{\beta}}$ from the same time delayed vectors,
\begin{equation}
\mathcal{Y}^{j}_t = \left [ \mathcal{O}^j_t,\mathcal{O}^j_{t-\tau},...,\mathcal{O}^j_{t-(p-1)\tau} \right],
\end{equation}
unless $\tau$ is sufficiently small to resolve $\mathcal{T_{\alpha}}$ and $p$ is sufficiently large to span $\mathcal{T_{\beta}}$. This can produce very high-dimensional time delay vectors that require extremely long and densely sampled observation trajectories for their construction and can introduce numerical challenges in embedding and reconstruction. Alternatively, we can construct separate time delay vectors from the same time series data with distinct time delays $\tau_{\alpha}$ and $\tau_{\beta}$ and concatenate these to construct a single time delay vector $\mathcal{Y}^{i;\alpha,\beta}_t$ from a time series in a single observable $i$, 
\begin{equation}
\begin{split}
    \mathcal{Y}^{i;\alpha,\beta}_t = [ &\mathcal{O}^i_t,\mathcal{O}^i_{t-\tau^{i;\alpha}},...,
    \mathcal{O}^i_{t-(p^{\alpha}-1)\tau^{i;\alpha}}, \\
    &\mathcal{O}^i_t,\mathcal{O}^i_{t-\tau^{i;\beta}},...,
    \mathcal{O}^i_{t-(p^{\beta}-1)\tau^{i,\beta} } ],
\end{split}
\end{equation}
where $p^{\alpha}$ and $p^{\beta}$ are the delay dimensionalities for each delay time. Consideration of dynamics occurring over multiple characteristic times allows for access to information about the evolution of a given dynamical system from a single time series, and presents a route to more accurate estimation of the multi-scale nature of $\mathcal{M}^{\prime}$.

\textbf{Choosing delay times and delay dimensionalities:} Takens' Theorem is silent on the particular choice of time delay, $\tau$ \cite{Takens}. In practical applications, however, the choice of $\tau$ can strongly affect the quality of the reconstruction. In general, $\tau$ should be tuned to match characteristic time scales of the dynamics we wish to reconstruct, should be large enough that successive coordinates contain independent information but not so large that they become dynamically unrelated, and should not correspond to an integer multiple of a dominant orbital period to avoid temporal aliasing \cite{kantz2004nonlinear,MI,martin2024robust}.
In the univariate case, delay times may be chosen by taking the first minimum of each observable's autocorrelation or mutual information function or by choosing a threshold value of the autocorrelation function \cite{MI}. For multi-temporal cases, we choose time delays that access different time scales, and this selection may be informed by \textit{a priori} physical knowledge of the system dynamics. 

In terms of the delay dimensionality, Takens' Theorem requires that the delay vectors possess a dimensionality of more than twice that of the intrinsic dimensionality of the system \cite{Takens}. Practically, we estimate the intrinsic dimensionality, $k$, from time series using the E1(d) method of Cao \cite{Cao97}, and select an embedding dimensionality of $p$ $\geq$ $(2k+1)$.

\section{\label{TARIII:sec:arch}Computational Application: Architecture of TAkens Reconstruction (TAR)}

\begin{figure*}[ht!]
    \centering
    \includegraphics[width=0.99\textwidth]{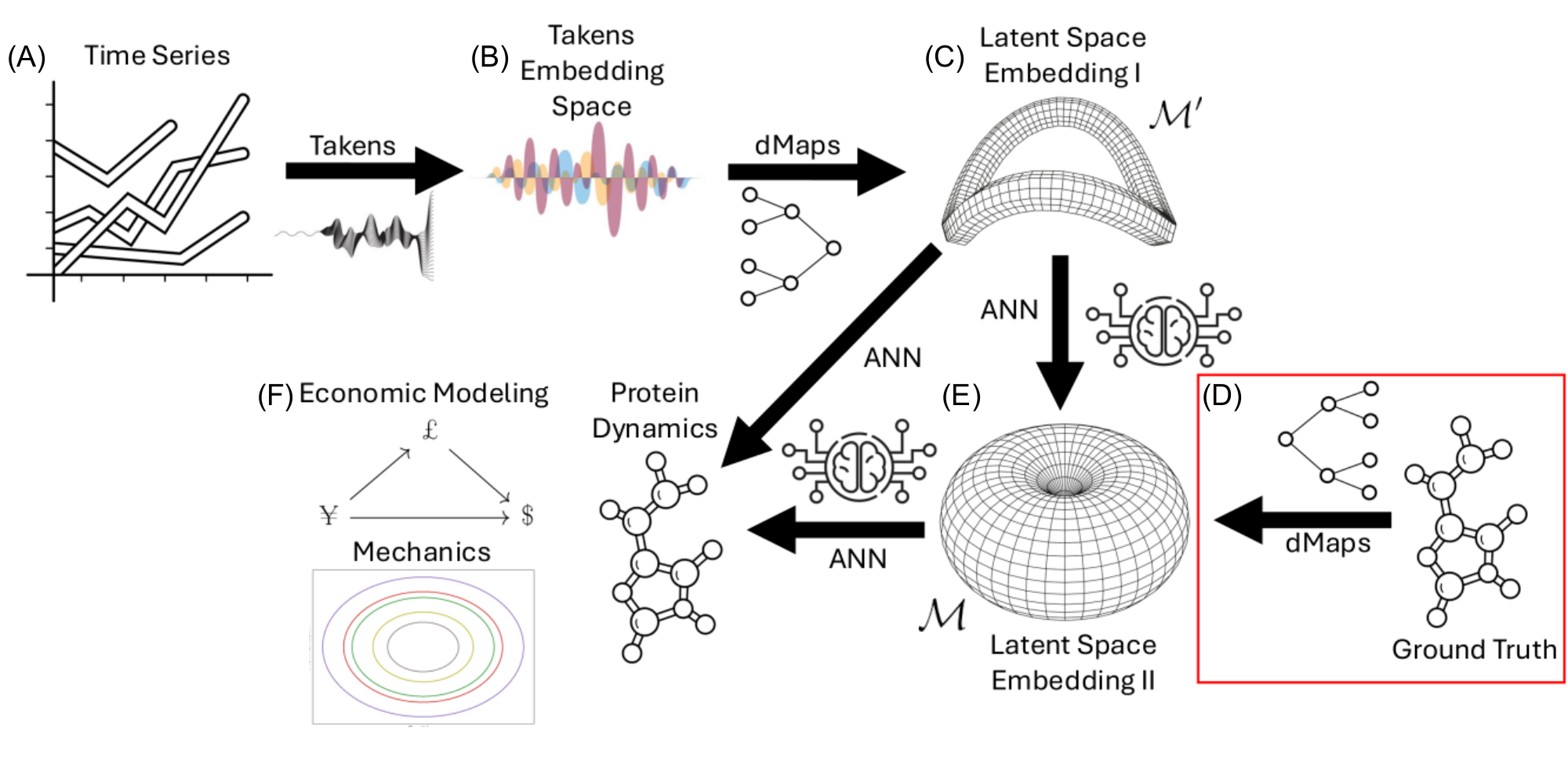}
    \caption{Schematic illustration of the TAkens' Reconstruction (TAR) approach. \textbf{(A)} The input to the TAR pipeline is a time series in one or more observables of a dynamical system. \textbf{(B)} The time series data are structured into either univariate, multivariate or multi-temporal Takens' delay vectors generating a Takens' delay embedding of the high-dimensional dynamical system. \textbf{(C)} Diffusion maps are employed to perform nonlinear manifold learning over the delay embedding vectors to learn an image of the intrinsic manifold, $\mathcal{M}^{\prime}$, of the dynamical system. \textbf{(D)} If full-dimensional training data is available, \textbf{(E)} diffusion maps manifold learning over the full-dimensional dynamical trajectories can be used to learn a the intrinsic manifold, $\mathcal{M}$. Takens' Delay Embedding Theorem asserts, under mild technical conditions, that the two learned manifolds $\mathcal{M}^{\prime}$ and $\mathcal{M}$ are related by a smooth, invertible and bijective mapping (i.e. diffeomorphism) that we approximate by an artificial neural network trained to map between the embedding of corresponding training points onto the two manifolds. \textbf{(F)} We train a second artificial neural network to perform a ``lifting'' operation from the manifold $\mathcal{M}$ back to the high-dimensional state vectors in the high-dimensional training data. Conceptually, this is the inverse of the diffusion map dimensionality reduction operation on these data that was used to construct $\mathcal{M}$. Once trained, the full TAR pipeline permits low-dimensional time series to be fed into the trained workflow and used to predictively reconstruct the high-dimensional system state through the path of operations \textbf{(A)} $\rightarrow$ \textbf{(B)} $\rightarrow$ \textbf{(C)} $\rightarrow$ \textbf{(E)} $\rightarrow$ \textbf{(F)}. The TAR approach can be applied to generic dynamical systems including molecular, mechanical, and economic systems.}
    \label{TARIII:fig:schematic}
\end{figure*}

The theoretical foundations detailed in Sections \ref{TARIII:sec:Dynamics} and \ref{TARIII:sec:delayvecs} provide the basis for a computational approach to reconstruct dynamical system states from low-dimensional time series. We term this approach TAkens' Reconstruction (TAR) and have developed an open-source software implementation to make this readily available to users interested in applying this formalism to arbitrary dynamical systems (Fig.\ \ref{TARIII:fig:schematic}). The TAR approach makes it possible to straightforwardly estimate the image $\mathcal{M}^{\prime}$ of the true intrinsic manifold $\mathcal{M}$ from low-dimensional time series via the application of diffusion maps nonlinear manifold learning to univariate, multi-variate, or multi-temporal delay embeddings and to compute distributions and paths over this topologically equivalent manifold. The projection of low-dimensional time series data onto $\mathcal{M}^{\prime}$ is already very useful since the topologically identical nature of $\mathcal{M}^{\prime}$ and $\mathcal{M}$ mean that the integrated probability over metastable states and the connectivity between them are preserved and one can gain deep understanding of relative probabilities and transition routes between metastable states \cite{Ferg16,Ferg18}.

If full-dimensional system observations are available, the TAR approach permits us to go further by direct estimation of $\mathcal{M}$ by the application of diffusion maps to these data, approximation of the diffeomorphic mapping from $\mathcal{M}^{\prime}$ to $\mathcal{M}$, and the approximate reconstruction (``lifting'') of the full-dimensional system state from its low-dimensional location on $\mathcal{M}$. The learned $\mathcal{M}^{\prime}$ to $\mathcal{M}$ mapping effectively reverses the artificial stretching and squashing induced on $\mathcal{M}^{\prime}$ relative to $\mathcal{M}$ by construction of the delay embeddings and so permits not just topological but also topographical interpretation of paths and distributions constructed over $\mathcal{M}$ \cite{topel2020,topel2023}. Specifically, probability distributions over the intrinsic manifold $\mathcal{M}$ faithfully reflect those over the reduced-dimensional projection of the full-dimensional system uncorrupted by the unknown diffeomorphic transformation induced over $\mathcal{M}^{\prime}$, such that the stationary distribution over the manifold and density of transition pathways between states are immediately interpretable and meaningful. These distributions, which can be empirically estimated by the projection of sufficiently long time series in the low-dimensional observables onto $\mathcal{M}^{\prime}$ and then transforming these over to $\mathcal{M}$, can be extremely valuable in exposing, for example, the relative probabilities of metastable states of a molecule or the various routes by which a ecological system oscillates between different metastable proportions of predators and prey. Moreover, the second learned mapping defining a lifting from $\mathcal{M}$ back to the full-dimensional system state provides mechanistic understanding of the various locations over the manifold to interpret the various metastable states and interconversion pathways. For example, two distinct pathways between the same pair of metastable states can be lifted back to the full-dimensional space to resolve the different mechanisms by which these two transitions occur. In a financial system, this could correspond to different temporal orderings of a change in stock prices in different market sectors, where the price change in the first sector leads that in the second in one pathway, whereas this ordering is temporally flipped in the second pathway.

In the following subsections, we proceed to describe each component of the TAR pipeline summarized above. We first assemble our time series input data into univariate, multivariate or multi-temporal Takens' vectors. We then reduce vector dimensionality with diffusion maps on to the intrinsic manifold. If high-dimensional training data is available, as is the case in analyzing numerical simulations where this ground truth is available (Sections \ref{TARIII:sec:toy}, \ref{TARIII:sec:Villin}), we learn approximate maps back to the full-dimensional phase space using artificial neural networks (ANNs). The trained TAR pipeline then presents a means to predictively reconstruct the high-dimensional system state from low-dimensional time series observations. In the absence of access to high-dimensional training data (Section \ref{TARIII:sec:VOO}), the learned low-dimensional manifold from the Takens' delay embedding vectors can provide a valuable means to understand and predict the dynamical evolution of the system. We have previously developed and relayed components of this protocol in the context of molecular systems in our  STAR (Single-molecule TAkens Reconstruction) pipeline \cite{topel2020,topel2023}. The present TAkens Reconstruction (TAR) work extends this approach to arbitrary dynamical systems. We demonstrate the applications of TAR to simulated and observed ecological (Section \ref{TARIII:sec:toy}), molecular (Section \ref{TARIII:sec:Villin}), and financial (Section \ref{TARIII:sec:VOO}) systems.

\subsection{\label{TARIII:sec:train} Construction of Delay Vectors}

The input to the TAR pipeline is a time series in one or more observables of a dynamical system (Fig.\ \ref{TARIII:fig:schematic}(A)). This may constitute a single long time series or multiple short discontinuous time series. In general, we assume that these data are collected at a uniform sampling period. These time series are then processed into delay vectors (Fig.\ \ref{TARIII:fig:schematic}(B)). In this work, we explore the construction of univariate and multi-temporal delay vectors, as detailed in Section \ref{TARIII:sec:delayvecs}. If the observables contain spurious symmetries that are not present in the full-dimensional system (e.g., an observable measuring the linear extent of a molecule cannot distinguish the head-to-tail orientation of the molecule), then the system may only be reconstructed up to those symmetries such that states related by these symmetric operations cannot be discriminated \cite{topel2020}.

\subsection{\label{TARIII:sec:red} Manifold Learning}

We employ diffusion maps \cite{coifman2006diffusion,coifman2008diffusion,ferguson2011cpl} to learn representations of the intrinsic manifold $\mathcal{M}^{\prime}$ from Takens' delay vectors (Fig.\ \ref{TARIII:fig:schematic}(C)) and, if available, $\mathcal{M}$ from full-dimensional system observations (Fig.\ \ref{TARIII:fig:schematic}(E)). Given a high-dimensional data set $V \in \mathbb{R}^N$, corresponding to delay vectors of low-dimensional observations or high-dimensional vectors representing instantaneous observations of the full-dimensional system, we apply diffusion maps following a standard procedure: we compute the pairwise distances between each point $v \in V$, construct a kernel matrix by applying a Gaussian convolution to the pairwise distance matrix, row normalize the resulting matrix to form a right stochastic transition matrix, perform a similarity transform to a symmetric matrix, and finally diagonalize this matrix to yield the eigenvectors constituting the spectral embedding of the data into a diffusion map embedding \cite{coifman2006diffusion,coifman2008diffusion,landmarkdiffusion,ferguson2010systematic}. In general, we adopt the Euclidean metric to measure similarity of the high-dimensional data points, but other distance metrics may be employed depending on the character of the system. We choose an appropriate Gaussian kernel width by recursively performing diffusion maps over a small training set and estimate the intrinsic dimensionality $k$ using an E1(d) analysis  \cite{Cao97,coifman2006diffusion}. The intrinsic dimensionality motivates a delay embedding dimensionality of $p = (2k+1)$ to properly embed the system without spurious intersections \cite{Takens}. The embedding vectors are processed with diffusion maps and an appropriate embedding dimensionality estimated by identifying a gap in the diffusion map eigenvalue spectrum using the L-method \cite{salvador2004determining} and taking the larger of that value and $k$. For large data sets, we perform diffusion maps on a subset of the data and the remaining points are projected into the resulting manifold using the Nystr\"{o}m extension \cite{nystrom,landmarkdiffusion,wang2018study}.

\subsection{\label{TARIII:sec:inter} Reconstruction }

Takens' Delay Embedding Theorem asserts, under mild technical conditions, that the two learned manifolds $\mathcal{M}^{\prime}$ and $\mathcal{M}$ are related by a smooth, invertible and bijective mapping (i.e. diffeomorphism) \cite{Takens}. In numerical simulations of dynamical systems, we typically have access to the high-dimensional training data (Fig.\ \ref{TARIII:fig:schematic}(D)) from which the low-dimensional time series were extracted (Fig.\ \ref{TARIII:fig:schematic}(A)). These data sets furnish both the $\mathcal{M}^{\prime}$ and $\mathcal{M}$ manifolds, respectively, but, by virtue of their temporal alignment, a one-to-one correspondence of delay vectors to the high-dimensional system state exists. For a $p = (2m+1)$-dimensional univariate delay vector containing a concatenation of observations from times $t$ to $(t-2m\tau)$, we construct this correspondence between the first element of the delay vector $\mathcal{O}_t$ and the corresponding instantaneous high-dimensional system configuration $\textbf{r}(t)$ from which this low-dimensional observation was drawn. We follow this same practice for multivariate delay vectors and apply this same logic of construction to multi-temporal delay vectors so that delay vectors share observations at time $t$. Takens' Theorem asserts only that this mapping exists, subject to symmetries, sufficient sampling, appropriate delay time, and sufficiently high delay dimensionality, but it can be numerically approximated by learning over the pairs of corresponding training points. We achieve this using simple fully-connected feedforward artificial neural networks (ANNs) to learn the mapping from points in the manifold $\mathcal{M}^{\prime}$ extracted from the delay embedding vectors to the manifold $\mathcal{M}$ extracted from full-dimensional trajectories (Fig.\ \ref{TARIII:fig:schematic}(C) $\rightarrow$ (E)). In cases where the full-dimensional trajectory is already low-dimensional we can dispense with $\mathcal{M}$ and learn the mapping directly to the state vectors of the dynamical system. 

The high-dimensional training data can also be used to learn a ``lifting'' from the manifold $\mathcal{M}$ back to the high-dimensional state vectors (Fig.\ \ref{TARIII:fig:schematic}(F)). Conceptually, this operation is the inverse of the dimensionality reduction operation accomplished by diffusion maps in learning $\mathcal{M}$ and can be used to approximately reconstruct the high-dimensional state of the system by learning to fill in the degrees of freedom lost under the low-dimensional embedding into $\mathcal{M}$ \cite{EQfree,kevrekidis2009equation,jones2023diamondback,jones2025flowback,sidky2020molecular}. It is conceivable that one could learn a lifting operation to the high-dimensional system state from the manifold $\mathcal{M}^{\prime}$ using a single ANN and without first passing through $\mathcal{M}$. Empirically, we find more robust and higher accuracy reconstructions using the two-step lifting via $\mathcal{M}$. For the systems studied in this work, simple fully-connected feedforward ANNs are sufficient to provide satisfactory reconstruction accuracy by training them on the one-to-one mapping to high-dimensional state vectors from their images on the manifold $\mathcal{M}$. For more complex systems, more sophisticated lifting architectures designed to respect the fundamental system symmetries and generalize to out-of-sample data may be warranted, including, for example, generative adversarial networks, denoising diffusion models, or conditional flow matching \cite{ jones2023diamondback,jones2025flowback,sidky2020molecular}. 

Once the two ANNs are trained, the TAR workflow permits low-dimensional time series to be fed into the trained workflow and used to predictively reconstruct the high-dimensional system state through the path of operations Fig.\ \ref{TARIII:fig:schematic}(A) $\rightarrow$ (B) $\rightarrow$ (C) $\rightarrow$ (E) $\rightarrow$ (F). In the absence of high-dimensional training data, the low-dimensional manifold $\mathcal{M}^{\prime}$ learned from the delay vectors still provides a valuable tool to understand the key features of the system dynamics, the topology and connectivity of the state space, and, predict the dynamical evolution of the system from a short history of system observations as is done in empirical dynamical modeling (EDM) \cite{PMID:25733874,Sugiharacritnet}.

\section{\label{TARIII:sec:toy}Time Delay Optimization: Lotka-Volterra Predator-Prey Dynamics}

As a first demonstration of TAR, we apply it to numerical simulations of the Lotka-Volterra model of predator-prey dynamics \cite{LV}. Here we show that TAR can reconstruct the full two-dimensional system state (i.e., the predator and prey populations) from univariate time series (i.e., observing just the predator population) using various time delays. We find that reconstruction qualities are dependent on time delay choice, but, at least for simple periodic systems like this one, we can use time delays significantly longer than the characteristic time scales of dynamics. Excessively short time delays, however, tend to produce poor reconstruction accuracies.

The Lotka-Volterra equations describe the interactions of a two-species ecosystem of predators and prey through the coupled nonlinear first order differential equations,
\begin{equation}
    \dfrac{dX}{dt}= \alpha X - \beta XY ;\hspace{1cm}
    \dfrac{dY}{dt}= \delta XY - \gamma Y
\end{equation}
where $X$ and $Y$ are, respectively, the instantaneous predator and prey populations, and $\alpha=1$, $\beta=0.005$, $\delta=0.8$ and $\gamma=0.6$. We propagate the Lotka-Volterra coupled population dynamics equations for 100,000 time units using a Runge-Kutta 4\textsuperscript{th} order \cite{Butcher_1964} ordinary differential equation solver with a time step of 0.001 time units. This yields a 2D trajectory recording the predator and prey populations at 100,000,000 discrete time points. In practice, we construct 100,000 delay vectors and use the first 5000 for TAR training and the remaining 95,000 for testing. Longer simulation times are required to account for long time delays which require $2m\tau$ time units to construct the first delay vector of a given Takens embedding. Further details of the simulation and application of TAR are provided in \blauw{Section S1} of the \blauw{Supporting Information}.

\begin{figure}[ht!]
    \centering
    \includegraphics[width=0.99\columnwidth]{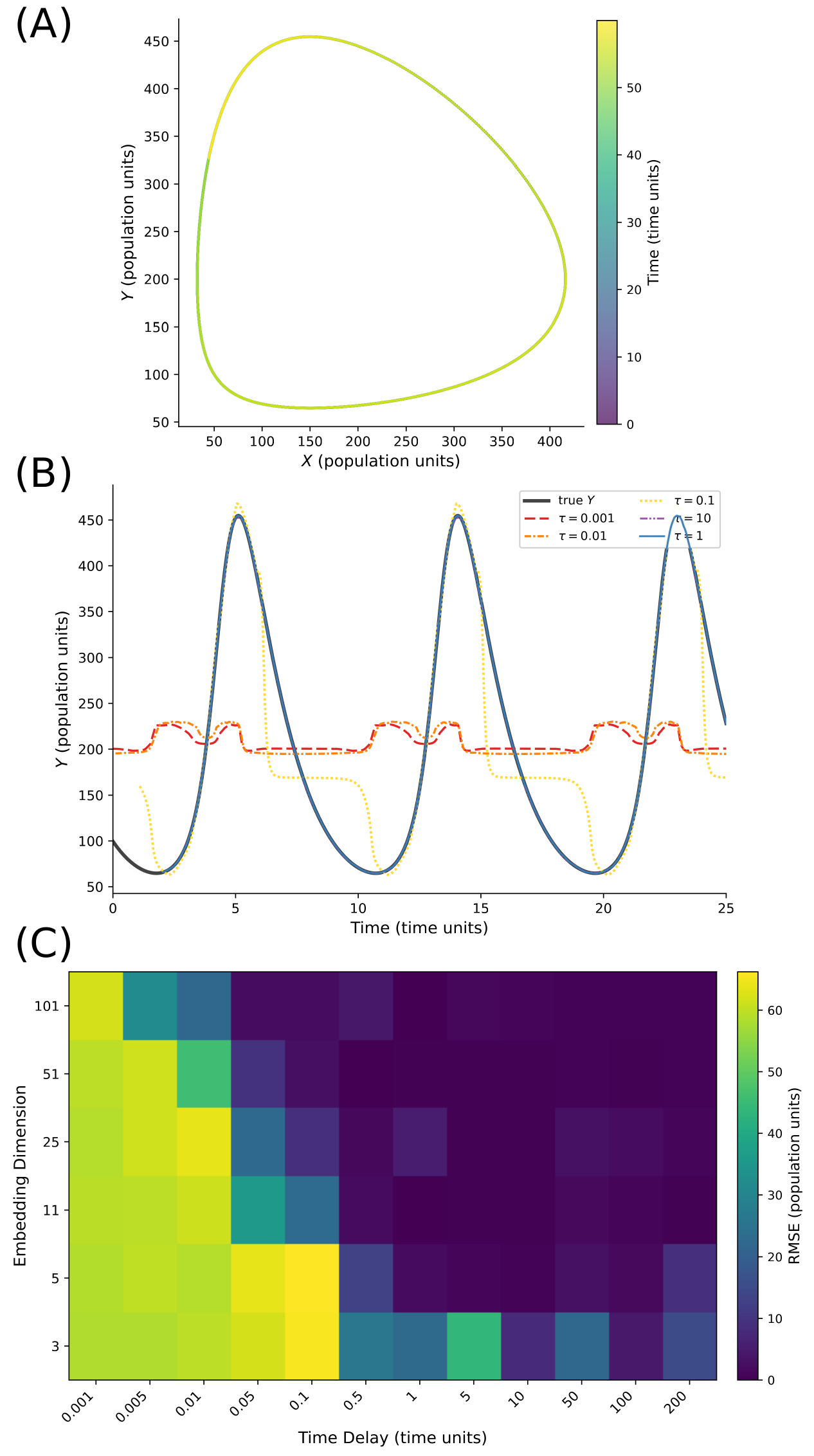}
    \caption{TAR reconstruction of the Lotka-Volterra predator-prey model. \textbf{(A)} Phase space diagram of the $(X,Y)$ limit cycle of predator $X$ and prey $Y$ populations. The period of the limit cycle is approximately 8.95 time units. \textbf{(B)} Reconstructed and true prey populations $Y$ over a 25 time unit period from the test set with delay times $\tau=[0.001,0.01,0.1,1,10]$ time units and delay vector dimensionality $p$ = 11 . \textbf{(C)} Root mean squared error (RMSE) reconstruction error over the test set as a function of delay time $\tau$ and embedding dimension $p$. As a baseline comparison, an RMSE $\approx (111.5 \pm 0.1)$ population units results from simply predicting the mean populations $(\bar{X},\bar{Y})$ at all test points.} 
    \label{TARIII:fig:LV}
\end{figure}

In Fig.\ \ref{TARIII:fig:LV}(A), we present a phase space diagram illustrating the limit cycle of the $(X,Y)$ populations over the course of our simulation. We adopt the predator population $X$ as our system observable from which to construct Takens' delay vector embeddings and attempt to reconstruct the full-dimensional system state, which, for this simple 2D system, amounts to predicting the prey population $Y$. For this choice of model parameters, the limit cycle possesses a period of approximately 8.95 time units.

In Fig. \ref{TARIII:fig:LV}(B), we present the resulting predicted $Y$ populations over the course of a sample 25 time unit period from the test set for the range of delay times $\tau=[0.001,0.01,0.1,1,10]$ time units and delay vector dimensionality $p$ = 11. We choose these time delays to sample many different orders of magnitude of dynamics while the embedding dimensionality $p$ = 11 far exceeds twice the $k$ = 1 intrinsic dimensionality of the limit cycle attractor, as required by Takens' Theorem. As observed in Section \ref{TARIII:sec:delayvecs}, empirical methods exist to estimate an appropriate delay time and embedding dimension for systems in which we lack \textit{a priori} knowledge. Reconstruction errors are determined by computing the root mean squared error (RMSE) between the $(X,Y)$ pair predicted by TAR and the actual values recorded within the test portion of the numerical simulation trajectory. Uncertainties in RMSE are computed from three-fold block averaging over the test partition. 

For a very short delay time $\tau = 0.001$ time units relative to the period of the system, the unobserved prey population is reconstructed poorly with RMSE = $(59.47 \pm 1.23)$ population units (red line), corresponding to a $\sim$30\% relative error. (Decomposing this into $X$ and $Y$ population RMSEs, we find $X$ is trivially reconstructed with very small error since the delay vector contains the predator population $X$ directly and so the error resides predominantly in reconstruction of the unobserved prey population $Y$.) The second shortest delay time of $\tau$ = 0.01 (orange line) performs scarcely any better, and only by the third shortest delay time $\tau$ = 0.1 (gold line) can the reconstruction begin to capture the broad trends in the $Y$ dynamics. For $\tau$ = 1 (blue line) and 10 (purple line), we achieve high quality reconstructions with RMSE = $(0.231\pm 0.008)$ and $(0.438\pm 0.145)$ population units, respectively, corresponding to relative errors of $\sim$0.12\% and $\sim$0.22\% respectively.

We present in Fig.\ \ref{TARIII:fig:LV}(C) the RMSE values resulting from a more comprehensive screen over various combinations of delay time $\tau$ = $[0.001,0.005,0.01,0.05,0.1.0.5,1,5,10,50,100,200]$ time units and delay embedding dimensions $p$ = $[3,5,11,25,51,101]$. The time span of the most compressed delay vector is just $p\tau$ = 0.003 time units while the most expansive delay vector spans $p\tau$ = 20,200 time units. The data exhibits a clear separatrix between a regime with poor reconstruction accuracy (lower left; yellow/green) and one with high accuracy (upper right; blue). The locus of this boundary is approximately defined by the conditions that TAR models with $p\tau$ $<$ 1 time unit perform reconstruction with comparable accuracies to just guessing the mean, whereas TAR models with $p\tau$ $>$ 5 time units (approximately half the period of the limit cycle) can perform reconstructions of the $Y$ with accuracies of 1 population unit or better, corresponding to a $\sim$0.5\% relative error.

Our analysis of the simple 2D Lotka-Volterra model as a periodic, non-chaotic, toy system exposes several notable behaviors that can provide intuition for applications to higher-dimensional and more complex dynamical systems. In addition to a delay embedding dimension exceeding twice the intrinsic dimensionality of the system (i.e., $p \geq (2k+1)$) as required by Takens' Theorem, the time spanned by the delay vector $p\tau$ must be sufficiently large for accurate reconstructions. Empirically, we observe a critical threshold of approximately half a period of the limit cycle. The common empirical techniques to tune the delay time by identifying the first minimum of the autocorrelation or mutual information functions \cite{MI} suggest time delays of 1-3 time units. However, we find delay vectors spanning approximately a half period or more are capable of accurate system reconstruction over a wide range of delay times and embedding dimensions.

\section{\label{TARIII:sec:Villin} Multi-Temporal Embedding: Molecular Dynamics Simulations of the Protein Villin}

In the previous section, we demonstrated how trading-off the time delay and delay embedding dimensionality can strongly influence the reconstruction accuracy and presented rules of thumb grounded in the empirical analysis of a simple toy system to guide their selection. In this section, we explore a multi-temporal approach to Takens' Theorem in which manifolds constructed using multiple time delays can offer a more complete description of dynamics than delay embeddings employing a single time scale. Information cascades may allow for the reconstruction of dynamics across time scales using a single delay time due to weak nonlinear couplings, feedback loops, multi-scale interactions, and stochastic resonances, permitting accurate reconstruction without such a multi-temporal approach \cite{Haken1983,networkcascade,kleinberg2010networks,Sugiharacritnet}. However, multi-temporal delay embeddings hold promise in more efficient and parsimonious reconstructions across time scales
and a primary motivation for multi-temporal embeddings of systems possessing multiple characteristic dynamical time scales is driven by computational pragmatism. To take one example of a canonical multi-scale system, dynamical motions in protein molecules can span several orders of magnitude, with bond stretching motions proceeding on $\mathcal{O}$(fs), rotomeric dynamics occur on $\mathcal{O}$(ps-ns), $\alpha$-helix and $\beta$-sheet formation on $\mathcal{O}$($\mu$s), and folding in excess of $\mathcal{O}$(ms) \cite{junghare2023markov,proteinTS}. A single-$\tau$ Takens' delay vector that spans all time scales must employ a short enough delay time $\tau$ to capture the fast fs bond vibrations dynamics would require a delay embedding dimensionality of $p = 10^{12}$ to also be responsive to the slow ms folding dynamics.
Constructing multi-temporal embeddings by concatenating delay vectors at different lag times is therefore critical for computational tractability in the analysis of multi-scale systems.

We demonstrate the multi-temporal approach in an application to the 125.6 $\mu$s MD simulation of the 35-residue mini-protein Villin produced by D.E.~Shaw Research and comprising 627,907 frames saved at 200 ps intervals \cite{deshaw}. We adopt as an observable the head-to-tail intramolecular distance between the first and last atoms of the linear protein chain extracted from the MD simulation trajectory. Conceptually, this univariate time series may be considered as an idealized and synthetic time-resolved observation of the molecular system that, in principle, is experimentally accessible by techniques such as single molecular F\"orster resonance energy transfer (smFRET) \cite{topel2020}. We use the first 80\% of the trajectory for training and the terminal 20\% as a hold-out test partition upon which we evaluate the reconstruction accuracy of the molecular state using TAR. Full details of the MD simulations are available in Lindorff-Larsen \textit{et al.}\ \cite{deshaw}.

The multi-temporal TAR approach requires a straightforward generalization of the workflow presented in Fig.~\ref{TARIII:fig:schematic}. Given a univariate time series in our system observable, we construct $k$ sets of time delay vectors with delay times $\tau_1,\tau_2...\tau_k$ and then learn the $k$ corresponding latent space manifolds $\mathcal{M}^{\prime}_1,\mathcal{M}^{\prime}_2...\mathcal{M}^{\prime}_k$ using nonlinear dimensionality reduction. There remains a single manifold $\mathcal{M}$ learned from the high-dimensional time series. Takens' Theorem asserts that each of the $\{\mathcal{M}^{\prime}_i\}_{i=1}^k$ manifolds are diffeomorphisms of $\mathcal{M}$ and so all, in principle, carry the same information on the dynamical evolution of the system. In practice, however, each $\mathcal{M}^{\prime}_i$ is more responsive to the dynamical evolution on the characteristic time scale corresponding to the delay time at which it was constructed, and we seek to obtain more robust reconstructions of $\mathcal{M}$ by integrating information over all $\{\mathcal{M}^{\prime}_i\}_{i=1}^k$. To do so, we train an ANN to map the concatenated vectors corresponding to the locations of a single training point on each of the $\{\mathcal{M}^{\prime}_i\}_{i=1}^k$ manifolds to the corresponding location on $\mathcal{M}$. In essence, we first construct a latent space at each delay time using standard TAR, then learn a multiplexed mapping from the union of these manifolds to the ground truth manifold learned from the high-dimensional training data. Finally, we train a second ANN to learn a lifting back to the high-dimensional system state from $\mathcal{M}$ in an identical manner to the single time delay case. Full details of the trajectory processing and application of the TAR workflow is provided in \blauw{Section S2} of the \blauw{Supporting Information}.

\begin{figure*}[ht!]
    \includegraphics[width=0.75\textwidth]{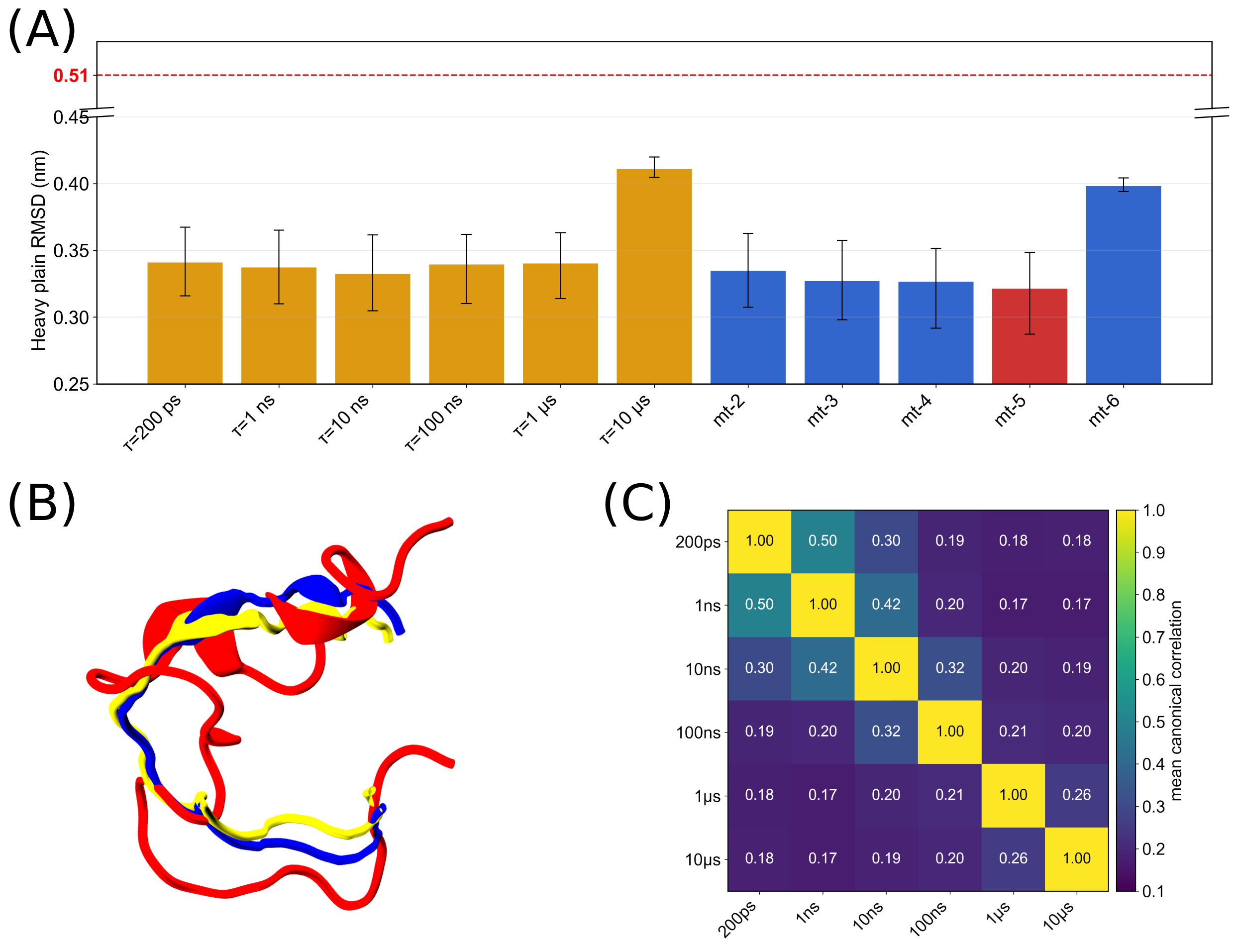}
    \caption{Multi-temporal TAR applied to the protein Villin. \textbf{(A)} Heavy-atom test RMSD reconstruction accuracies over the 20\% hold out test partition of the MD simulation trajectory under single-temporal (orange bars) and multi-temporal (blue bars) TAR. For single-temporal TAR, the delay time $\tau$ is reported under the bar and spans $\tau$ = [200 ps, 1 ns, 10 ns, 100 ns, 1 $\mu$s, 10 $\mu$s]. For multi-temporal TAR, mt-$k$ denotes a cumulative incorporation of successively longer single-temporal delay times, \textit{viz.} mt-2: [$\tau_1$ = 200 ps, $\tau_2$ = 1 ns], mt-3: [$\tau_1$ = 200 ps, $\tau_2$ = 1 ns, $\tau_3$ = 10 ns], mt-4: [$\tau_1$ = 200 ps, $\tau_2$ = 1 ns, $\tau_3$ = 10 ns, $\tau_4$ = 100 ns], mt-5: [$\tau_1$ = 200 ps, $\tau_2$ = 1 ns, $\tau_3$ = 10 ns, $\tau_4$ = 100 ns, $\tau_5$ = 1 $\mu$s], and mt-6: [$\tau_1$ = 200 ps, $\tau_2$ = 1 ns, $\tau_3$ = 10 ns, $\tau_4$ = 100 ns, $\tau_5$ = 1 $\mu$s, $\tau_6$ = 10 $\mu$s]. The best reconstruction accuracy is achieved by mt-5 with a heavy-atom RMSD of 0.321 nm and is indicated as a red bar. Uncertainties in reconstruction accuracy are denoted by error bars computed by bootstrap resampling and denote 95\% confidence intervals. The heavy-atom RMSD reconstruction performance of 0.51 nm achieved by guessing a constant molecular configuration corresponding to the training set configuration with the lowest RMSD to all other training set configurations is denoted by a dashed red line. \textbf{(B)} Molecular rendering of a representative test-set Villin structure showing the all-atom ground-truth structure (red) overlaid with a TAR reconstructions from the best performing single-temporal TAR model with $\tau$ = 10 ns (heavy-atom RMSD = 0.332 nm) (yellow) and the best performing multi-temporal TAR model mt-5 (heavy-atom RMSD = 0.321 nm) (blue). Molecular renderings produced with VMD \cite{Schulten:96:JMG}. \textbf{(C)} Pairwise canonical correlation analysis (CCA) over the diffusion map embeddings defining the $\{\mathcal{M}^{\prime}_i\}_{i=1}^k$ manifolds constructed at delay times $\{\tau_i\}_{i=1}^k$. The matrix reports the mean canonical correlation coefficient taken over the CCA-transformed canonical variables calculated over the leading six dimensions of the diffusion map embedding defining each manifold $\mathcal{M}^{\prime}_i$ constructed at delay time $\tau_i$. Values are restricted to lie on the interval $[0,1]$, where a value of unity indicates perfect linear correlation under the CCA.}
    \label{TARIII:fig:Villin}
\end{figure*}

In Fig.~\ref{TARIII:fig:Villin}(A) we compare the heavy-atom RMSD reconstruction accuracies over the 20\% hold out test partition of the MD simulation trajectory under single-temporal and multi-temporal TAR. As a baseline comparison for the multi-temporal models, we report the performance of single-temporal TAR at delay times $\tau$ = [200 ps, 1 ns, 10 ns, 100 ns, 1 $\mu$s, 10 $\mu$s] (orange bars). In all cases, delay embedding dimensionalities of $p$ = 11 were employed, which are in excess of twice the 3D intrinsic dimensionality of the Villin intrinsic manifold \cite{topel2023}. We compare the performance against multi-temporal TAR constructed by incorporating successively longer delay times into the reconstruction (blue bars), which are denoted mt-$k$, indicating a multi-temporal embedding incorporating the $k$ shortest single-temporal delay times, \textit{viz.} mt-2: [$\tau_1$ = 200 ps, $\tau_2$ = 1 ns], mt-3: [$\tau_1$ = 200 ps, $\tau_2$ = 1 ns, $\tau_3$ = 10 ns], mt-4: [$\tau_1$ = 200 ps, $\tau_2$ = 1 ns, $\tau_3$ = 10 ns, $\tau_4$ = 100 ns], mt-5: [$\tau_1$ = 200 ps, $\tau_2$ = 1 ns, $\tau_3$ = 10 ns, $\tau_4$ = 100 ns, $\tau_5$ = 1 $\mu$s], and mt-6: [$\tau_1$ = 200 ps, $\tau_2$ = 1 ns, $\tau_3$ = 10 ns, $\tau_4$ = 100 ns, $\tau_5$ = 1 $\mu$s, $\tau_6$ = 10 $\mu$s]. Again, delay embedding dimensionalities of $p$ = 11 were employed for each individual delay time. Uncertainties are reported as 95\% confidence intervals computed by a moving-block bootstrap with 1000 resamples and 1 $\mu$s (5000 frame) blocks selected to exceed the longest dynamical autocorrelation time in the MD trajectories. We also report the null case heavy-atom RMSD reconstruction accuracy of 0.51 nm that is achieved by selecting the single training-set configuration whose mean RMSD against all other training configurations is smallest and predicting this configuration as a the reconstruction at every test set instant (red dashed line). In Fig.~\ref{TARIII:fig:Villin}(B) present an illustrative molecular visualization overlaying a selected ground-truth MD configuration with reconstructions predicted from the top performing single-temporal and multi-temporal TAR models.

The single-temporal TAR reconstructions achieve approximately 0.35 nm heavy-atom RMSD reconstruction accuracies, with the $\tau$ = 10 ns delay time providing the highest accuracy at 0.332 nm. The $\tau$ = 10 ns delay time corresponds to the first  minimum of the autocorrelation function of the head-to-tail observable, providing empirical support for the selection of optimal delay times based on the autocorrelation curve. Nevertheless, the RMSDs for time delays spanning 200 ps to 1 $\mu$s all lie within 95\% confidence intervals, indicating statistically similar performance. We note that the large degradation in accuracy for the $\tau$ = 10 $\mu$s delay time to a heavy-atom RMSD of 0.411 nm may be primarily attributed to this large delay time reducing the volume of useable training data, since, for an $p$ = 11-dimensional delay embedding dimensionality, the terminal $(p-1) \times \tau$ = 100 $\mu$s of the training trajectory cannot be used to construct delay embedding vectors. In all cases the TAR reconstructions are $\sim$30\% better than the 0.51 nm heavy-atom RMSD achieved by simply adopting the single best training configuration as the reconstruction of every test instant, demonstrating that the TAR pipeline is extracting and using information from the learned intrinsic manifold to make its predictions.  

Turning to the multi-temporal TAR reconstructions, we observe a monotonic increase in reconstruction accuracy as cumulatively more delay times are included in the multi-temporal workflow from mt-2 through mt-5, with mt-5, which incorporates the five delay times [$\tau_1$ = 200 ps, $\tau_2$ = 1 ns, $\tau_3$ = 10 ns, $\tau_4$ = 100 ns, $\tau_5$ = 1 $\mu$s], achieving the best reconstruction accuracy with a heavy-atom RMSD of 0.321 nm. This corresponds to a $\sim$3\% improvement in reconstruction accuracy over the best single-temporal $\tau$ = 10 ns TAR model, demonstrating that the incorporation of multiple delay times can improve TAR performance beyond what is possible from single-temporal models. We note, however, that the relatively small size of the MD trajectory test set means that this relatively small performance improvement does lie within the 95\% confidence intervals. We note that a significant drop in the reconstruction accuracy of mt-6 relative to mt-5 corresponding to a jump in the heavy-atom RMSD to 0.398 nm. The mt-6 model differs from mt-5 in that it incorporates a 6\textsuperscript{th} delay time of $\tau_6$ = 10 $ \mu$s for which we recall that the single-temporal TAR performed quite poorly due to a large reduction in the training data volume at this long delay time. This observation indicates that, contrary to what might have been anticipated, the incorporation of additional delay times with poor single-temporal performance can actually degrade the reconstruction accuracy of a high-performing multi-temporal model. The current multi-temporal TAR implementation treats all manifolds $\{\mathcal{M}^{\prime}_i\}_{i=1}^k$ equally by a simple concatenation of the training data locations over the $k$ manifolds in learning the mapping to $\mathcal{M}$. We propose that the introduction of a poorly learned manifold can disrupt learning of this mapping through the trained ANN and propose that more sophisticated mapping approaches may be considered that weight the $\mathcal{M}^{\prime}_i$ by, for example, training data volume.   


To quantify the complementarity of the information added by incorporating each additional delay time within the multi-temporal approach, we performed pairwise canonical correlation analysis (CCA) \cite{hotelling1936} over the diffusion map embeddings defining the $\{\mathcal{M}^{\prime}_i\}_{i=1}^k$ manifolds. Specifically, given a pair of manifolds $\mathcal{M}^\prime_i$ and $\mathcal{M}^\prime_j$ constructed at delay times $\tau_i$ and $\tau_j$, we conducted CCA over the leading six dimensions of the diffusion map embedding defining the manifold and report the mean canonical correlation coefficient, which is restricted to lie on the interval $[0,1]$, over the six CCA-transformed canonical variables. Mathematically, CCA identifies linear combinations of the embedding dimensions defining the manifolds to produce canonical variables with maximal mutual correlation. High mean canonical correlation may, therefore, be interpreted as the two manifolds sharing a high degree of linearly-accessible information (coinciding with the mutual information under a joint-Gaussian approximation), whereas low mean canonical correlation indicates that the two manifolds carry largely complementary information, at least at the level of a linear dependence. We report the results of the CCA analysis as a symmetric pairwise comparison matrix in Fig.~\ref{TARIII:fig:Villin}(C).

As expected, the main diagonal of the pairwise comparison matrix is populated by unit entries since comparing each manifold to itself should result in perfect canonical correlation. Also as expected, manifolds constructed at similar time delays tend to carry higher degrees of mean canonical correlation than those with more discrepant time delays. This is manifested by relatively large mean canonical correlation coefficients in the superdiagonal and subdiagonal, and lower values away from the main diagonal. The mean off-diagonal mean canonical correlation is 0.246, with the largest correlation 0.495 between the two shortest delays (200 ps $\leftrightarrow$ 1 ns) and the smallest 0.170 between the most temporally separated pair (1 ns $\leftrightarrow\ 10 \ \mu$s). A simple linear regression of the pairwise mean canonical correlations on the log temporal separation $\log_{10}|\tau_i - \tau_j|$ yields a strongly significant negative slope (Pearson $r = -0.81$, $p = 2.9\times 10^{-4}$), confirming that more temporally separated $\tau$ pairs contribute more independent information to the multi-temporal stack. 

The mean canonical correlation of 0.21 between the $\tau_4$ = 100 ns and $\tau_5$ = 1 $\mu$s delay times corresponds to the lowest entry in the matrix superdiagonal/subdiagonal, indicating that of all pairs of neighboring delay times, these two share the lowest mutual information content. The single largest stepwise improvement in the multi-temporal TAR analysis occurs between mt-4 and mt-5 upon incorporation of the $\tau_5$ = 1 $\mu$s time delay. We propose that this may be understood as this $\mathcal{M}^\prime_5$ manifold containing substantial new information on the system dynamics relative to $\{\mathcal{M}^\prime_i\}_{i=1}^4$. Interestingly, the characteristic folding and unfolding times of Villin in these trajectories are 2.8 $\mu$s and 0.9 $\mu$s, respectively, which are commensurate with the $\tau_5$ = 1 $\mu$s time delay, suggesting that this manifold contains dynamical information on global folding that is not contained in those constructed at shorter delay times.


Taken together, these results demonstrate the utility of multi-temporal Takens' delay embeddings in squeezing more information from a single univariate time series to modestly improve reconstruction accuracy in multi-scale systems by incorporating information across multiple time scales. Importantly, the multi-temporal approach does not entail an increase in the required number of observables as is necessitated by multivariate approaches \cite{Cao1998}, but the multivariate and multi-temporal approaches can be straightforwardly combined by constructing delay embeddings from each time series at different delay times. This approach is expected to be most efficacious when different observables are responsive to multiple different characteristic dynamical time scales. The observed lack of monotonic improvement with multi-temporal approaches demonstrates that time scale choice can be important to improve reconstruction accuracy, and highlights the importance of calibrating the delay time spectrum to the characteristic time scales of the system, either by trial-and-improvement, prior knowledge, or a spectral analysis of the time series data, and cautions against including low-quality data that can degrade reconstruction accuracy.

\section{\label{TARIII:sec:VOO} Phase Space Reconstruction in the Absence of Full-Dimensional Observations: Vanguard S\&P 500 ETF}

Many experiments describe phenomena that are beyond the capacity of simulation, either because the rule sets describing them are unknown or are excessively expensive. 
In such cases, it may not be possible to either access the full-dimensional state of the natural system or generate realistic synthetic trajectories by numerical simulation. 
Even in the absence of ground truth knowledge of the full-dimensional system the learned low-dimensional manifold from the Takens' delay embedding vectors can provide a useful means to view, understand, and predict the dynamical evolution of the system and gain insight into phenomenologies driving the only partially observed dynamics. Within the context of the TAR framework presented in Fig.~\ref{TARIII:fig:schematic}, this amounts to executing the first three steps (i.e., Fig.~\ref{TARIII:fig:schematic}A-C) to learn the latent embedding $\mathcal{M}^{\prime}$ from the Takens' delay embedding vectors. As a diffeomorphism of the true intrinsic manifold $\mathcal{M}$, the distributions over this embedding are modulated by an unknown invertible transformation, rendering the quantitative interpretation of relative stabilities (e.g., well-depths and barrier heights) over the stationary distribution subject to an unknown stretching and squashing. The preservation of the continuity and connectivity of the manifold under the guaranteed topological equivalence means that analysis of the projected dynamics can still reveal a wealth of information on transitions between metastable system states, excursions of the system into unstable regimes, oscillatory and periodic behaviors, and the response of the system to external perturbations, shocks, or changes in the prevailing conditions. We have previously exploited this relationship to demonstrate how to gain understanding of the stable states and transition pathways in molecular systems without access to full-dimensional observations \cite{Ferg16}. 

We demonstrate the application of TAR in the absence of ground-truth full-dimensional system observations in the analysis of a real-world economic data set comprising the execution price of shares of the Vanguard S\&P 500 Exchange Traded Fund (VOO) over the month of January 2024. These data can be viewed as a univariate time series serving as a low-dimensional observable of the dynamical evolution of the global economy. Obtaining a full-dimensional observation of the global economy or generating a realistic full-dimensional dynamical model are both intractable due to the scale, complexity, and unknown rules governing its dynamical evolution. Nonetheless, Takens' Theorem provides a lens through which to analyze these data and extract understanding of the dynamical evolution of the system. Specifically, we anticipate that the projected dynamics on the learned $\mathcal{M}^{\prime}$ manifold can expose these qualitatively different behaviors, present a means to identify regime shifts, and resolve anomalous excursions into new regions of phase space corresponding to previously unexplored dynamical regimes. These data also represent a good test case since the expected behaviors of this observable at various characteristic time scales are well understood -- mean reversion at short time scales, random walk at intermediate time scales, and trend following at long time scales -- allowing us to test the efficacy of TAR by its ability to expose and recapitulate these trends. 

At \textit{short time scales} of $\mathcal{O}$(ms-s) the returns distributions of equities are known to be mean reverting as a result of combined effects of bid-ask bounce, liquidity effects and associated transaction costs, order-book dynamics, and inventory management by market makers \cite{MADHAVANmicrostructure,Bouchaud_TQP,econometricsfinmark,OharaMicro}. While the canonical microstructure literature approaches the first-order autocorrelation in tick-level returns through Roll's bid-ask bounce model \cite{roll1984spread,econometricsfinmark,OharaMicro}, the marginal returns distribution at sub-minute scales is sharply peaked and heavy-tailed. For a more flexible approach to capture the marginal returns at these scales, one may apply the Normal-Inverse-Gaussian (NIG) distribution \cite{barndorffnielsen1997nig}, a four-parameter infinitely-divisible heavy-tailed distribution \cite{barndorffnielsen1997nig} used in the literature to approach higher order and alternate sourced microstructure phenomenologies which emerge in returns distributions from subordinated stochastic processes  \cite{clark1973time}. The probability density function is given by,
\begin{align}
\label{return:NIG}
    f(x;\alpha,\beta,\delta,\mu) &= \frac{\alpha\,\delta\,e^{\delta\gamma}}{\pi}\,\frac{K_1\!\bigl(\alpha\sqrt{\delta^2 + (x-\mu)^2}\bigr)}{\sqrt{\delta^2 + (x-\mu)^2}}\,e^{\beta(x-\mu)}, \notag \\ 
    \gamma &= \sqrt{\alpha^2-\beta^2},
\end{align}
where $\alpha>0$ and $0\leq|\beta|<\alpha$ control tail heaviness and skewness respectively, $\delta>0$ is a scale parameter, $\mu$ is the center of the distribution, and $K_1$ is the modified Bessel function of the second kind of order one. 

At \textit{intermediate time scales} of $\mathcal{O}$(s-min), the price of liquid assets, of which VOO is one example, is expected to follow a random walk since traders have similar information and consequently have no ability to make decisions based on material changes in market conditions \cite{RWFama,fama1965behavior}. The returns attributable to these random walks can be modeled as Gaussian distributions,
\begin{equation}
\label{return:Gaussian}
    f(x) = \frac{1}{\sqrt{2\pi\hat{\sigma}^2}} \exp\left(-\frac{(x - \hat{\mu})^2}{2\hat{\sigma}^2}\right)
\end{equation}
where $x$ are the observed returns, $\hat{\mu}$ is the estimated mean return and $\hat{\sigma}$ is the estimated standard deviation. In the symmetric limit ($\beta = 0$ with $\alpha,\delta\to\infty$ and $\delta/\alpha = \sigma^2$), NIG converges to $\mathcal{N}(\mu,\sigma^2)$ allowing for modeling both of the fat-tailed microstructure phenomena and random walks of longer time scales.

On \textit{long time scales} of $\mathcal{O}$(min-h), returns spectra, depending on the particular asset, can exhibit volatility clustering. Shocks such as information entering a market are followed by elevated variance producing time-varying conditional second moments \cite{mandelbrot1963variation,fama1965behavior}. The ARCH/GARCH family \cite{engle1982arch,bollerslev1986garch} of models was designed to account for such moments, with the GARCH(1,1) model using a Student-$t$ innovation distribution to capture fat tails \cite{bollerslev1986garch,bollerslev1987tgarch} serving as the canonical conditional-volatility benchmark in equity markets,
\begin{align}
\label{return:GARCH}
    x_t &= \mu + \sigma_t z_t,\quad z_t \sim t_\nu^{*}(0,1) \notag \\
    \sigma_t^2 &= \omega + \alpha_1\,(x_{t-1}-\mu)^2 + \beta_1\,\sigma_{t-1}^2
\end{align}
where $z_t$ is a Student-$t$ random variable standardized so that $\mathrm{Var}(z_t)=1$ (requires $\nu>2$), $\omega>0$, $\alpha_1,\beta_1\geq 0$, and $\alpha_1+\beta_1<1$ are the GARCH coefficients ensuring positivity and stationarity of the conditional variance, $\nu$, the Student-$t$ degrees-of-freedom parameter, controls the tail heaviness, and $\sigma_t^2$ allows for time variation via conditional variance ARMA(1,1) in the squared returns. 

At \textit{very long time scales} of $\mathcal{O}$(h-days), we expect to see shifting equilibria and the emergence of long-term trends in the returns as a result of macroeconomic or product specific forces. For example, qualitatively different return dynamics were induced by the collapse of Lehman Brothers in September 2008 followed by the global financial crisis and associated elevated volatility, the March 2020 COVID-19 onset and multi-year yield compression \cite{baker2020covid}, and the 2022 Federal Reserve rate hikes. Shorter-term shocks such as supply shocks or commodity price swings can result in faster price changes as new information enters the market. These long-term regime shifts are not expected to admit a single closed-form parametric description and their non-stationarity is the central feature motivating a more dynamic data-driven representation of these systems.

The January 2024 execution price of shares of the Vanguard S\&P 500 Exchange Traded Fund (VOO) were obtained from Wharton Research Data Services (WRDS) \cite{WRDS}, and constitute a univariate time series of 477,000 instantaneous share prices recorded over the 6.5 hours of trading each weekday at tick-level resolution. We choose to work with log returns as they better approximate a normal distribution, are additive in time, and are more stationary than standard returns \cite{econometricsfinmark}. We construct delay vectors with delay times of $\tau$ = 2 s, 1 min, and 1 h, as characteristic time scales expected to fall, respectively, into the mean reverting, random walk and trend following regimes. In constructing the delay vectors, we first coarsen trade ticks onto a uniform 1~s grid by retaining the last execution price and forward-filling any empty bins. In practice, every bin contains multiple ticks at the second-level resolution. Empty bins are common at shorter time scales and on less liquid assets and are conventionally handled by last-observation-carried-forward or midpoint quotes \cite{Bouchaud_TQP,OharaMicro}. We choose not to construct delay vectors spanning different trading days, but techniques exist to smooth over inter-day effects \cite{RVHansenLunde2006}. Delay embedding dimensionalities were estimated using the E1(d) method \cite{Cao97} to inform the use of $p$ = 3 delay embedding dimensionalities and the $\mathcal{M}^\prime$ intrinsic manifold was learned from the ensemble of delay vectors using diffusion maps. We conduct manifold learning over up to 5000 delay vectors selected randomly from the first half of the constructed delay vectors, corresponding approximately to the first half of January 2024 trading days. The remaining delay vectors are projected onto the $\mathcal{M}^\prime$ manifold using the Nystr\"{o}m extension \cite{nystrom}. Finally, an ANN to learn a mapping from the attractor $\mathcal{M}^\prime$ to the returns distribution. We note that since we do not have access to full-dimensional system observations or the true intrinsic manifold $\mathcal{M}$, this is not a lifting approximation to the full-dimensional space as in the previous sections of this work, but rather a backmapping of locations on $\mathcal{M}^\prime$ to the returns distribution that allows us to take observations of novel returns and map them to the returns distribution via the TAR framework. We elect to investigate this more stable object of returns distribution over individual returns as the stochastic nature of returns obscures individual returns trajectories while the statistical quantity of returns distributions remains stable over the studied periods. Full details of the data and analysis are provided in \blauw{Section S3} of the \blauw{Supporting Information}. 

In the absence of ground truth training data from which to train a model to learn the transformation from $\mathcal{M}^\prime$ to $\mathcal{M}$ and perform approximate full-dimensional reconstructions (c.f.\ Fig.~\ref{TARIII:fig:schematic}D-F), we assess TAR performance in two ways. First, we quantify the agreement between the empirical test-set return distribution and its TAR reconstruction, as measured by the total-variation (TV) distance in return space. The TAR reconstruction is obtained by mapping locations on $\mathcal{M}^\prime$ back to the returns distribution via the learned ANN backmapping $\psi \to r$ and the TV distance is evaluated between return distributions. The L1 norm between two discrete probability distributions $p$ and $q$ over a common bin set $\{i\}$ is $\|p-q\|_1 = \sum_i |p_i - q_i|$, twice their total-variation distance $\|p-q\|_\text{TV}$. A small TV distance indicates that the reconstruction is representative of the test data. Second, we benchmark TAR against the three parametric models  (NIG, Gaussian, and GARCH(1,1)-$t$) by  fitting the parameters $(\hat\alpha, \hat\beta, \hat\delta, \hat\mu)$ of the NIG model (Eqn.~\ref{return:NIG}), $(\hat\mu, \hat\sigma)$ of the Gaussian model (Eqn.~\ref{return:Gaussian}), and $(\hat\omega, \hat\alpha_1, \hat\beta_1, \hat\nu)$ of the GARCH(1,1)-$t$ model (Eqn.~\ref{return:GARCH}) on the training-half VOO return series.  We compare the learned distributions for NIG and Gaussian learned from all of January 2024 with the TAR model trained only on the first half of the data. For TAR and GARCH(1,1)-$t$, we only train on the first half of data. The asymmetry is in the prediction object: NIG and Gaussian are i.i.d.\ marginals, so a pooled fit only sharpens fixed parameters and, if anything, favors the benchmarks, whereas GARCH(1,1)-$t$ forecasts the one-step-ahead density anchored at $\sigma_T$, which a pooled fit would render self-referential.


\begin{figure*}[ht!]
    \centering
    \includegraphics[width=0.9\textwidth]{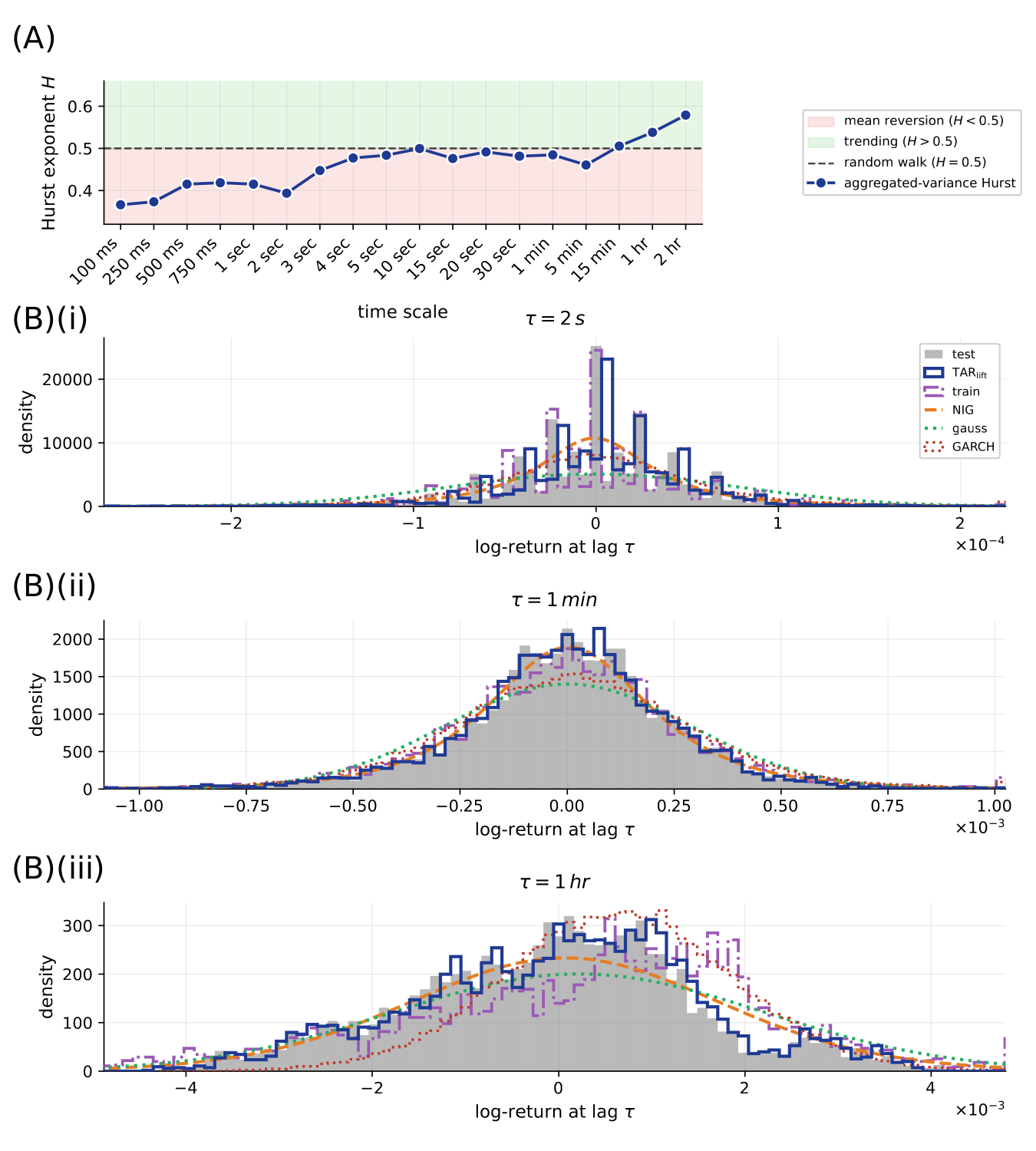}
    \caption{TAR analysis of the Vanguard S\&P 500 ETF (VOO) returns from January 2024. \textbf{(A)} Hurst exponents estimated across the time-scale ladder via the aggregated-variance method \cite{taqqu1995estimators}, exposing the three expected regimes: mean-reversion on millisecond to second time scales (0 < $H$ < 0.5), random walk on second to minute time scales ($H$ $\approx$ 0.5), and trending on hour time scales (0.5 < $H$ < 1).  \textbf{(B)} Empirical-test log-return distributions (gray fill) overlaid with five reconstructions at delay times of \textbf{(i)} $\tau$ = 2 s, \textbf{(ii)} $\tau$ = 1 min, and \textbf{(iii)} $\tau$ = 1 hr, corresponding to the mean reverting, random walk and trend following returns structures. We overlay the data with predictions from $TAR_{lift}$ reconstructions (blue), training-half empirical histogram (purple, ``train''), Normal-Inverse-Gaussian fit (orange, ``NIG''), Gaussian fit (green, ``gauss''), and GARCH(1,1)-$t$ simulation (red, ``GARCH''). Total-variation distances against empirical-test distribution corresponding to returns over the second half of January 2024 are reported in Table~\ref{tab:L1}. 
    }
    \label{TARIII:fig:VOO}
\end{figure*}

In Fig.\ \ref{TARIII:fig:VOO}(A), we present Hurst exponents calculated over the full January 2024 VOO time series across 100~ms to 2~hour time intervals using the aggregated-variance estimator \cite{taqqu1995estimators}.  The Hurst exponent $H \in [0,1]$ measures the degree to which autocorrelations in the time series decay as the lag time between points increases and quantifies the degree of long-term memory in the time series \cite{hurst1956p1,hurst1957p2,weron2002hurst}. Hurst exponents of 0 < $H$ < 0.5 are indicative of anti-persistent, mean reversion behavior, values of 0.5 < $H$ < 1 indicate persistent, trending behavior, and values of $H$ $\approx$ 0.5 correspond to memoryless, random walk behavior. On short time scales of milliseconds to seconds, we observe mean-reversion behavior with $H \approx 0.40$. At intermediate time scales of tens of seconds to several minutes, we observe random walk behavior with $H \approx 0.50$. At longer time scales hours, $H$ enters the trending regime, with $H$ $\approx$ 0.54 at 1~hr and 0.58 at 2~hr. 

In Fig.\ \ref{TARIII:fig:VOO}(B), we present histograms of the observed returns over $\tau=2$~s, $\tau=1$~min, and $\tau=1$~h time periods and the distributions generated by the best-fit NIG, Gaussian, and GARCH(1,1)-$t$ models on the training half. We quantify the quality of the fits to the test data in Table \ref{tab:L1}, where we report the total-variation (TV) distances between empirical-test return distributions and candidate reconstructions for each delay time. All results are reported in a return space on a shared normalized support across the test, $TAR_{lift}$, NIG, Gaussian, and GARCH(1,1)-$t$ distributions. The TAR column reports the TV distance between test returns and the ANN lift-reconstruction $\psi \to r$ trained on the first half of the data and forward-evaluated on $\psi$-test, while the NIG, Gauss and GARCH(1,1)-$t$ columns report the analogous TV against parametric fits on all data for the three non TAR parametric fit models. This is, in a sense, a harder test for TAR. It is trained on the first half and has to predict returns across the test half time series based on observed delay vectors.

\begin{table}[htbp]
    \centering
    \caption{Total-variation (TV) distances between empirical-test and four candidate reconstructions of VOO log-returns from January 2024 at delay times $\tau \in \{2\,\text{s}, 1\,\text{min}, 1\,\text{h}\}$ corresponding to mean-reverting, random-walk, and trend-following regimes. NIG and Gaussian parameters are fit by maximum likelihood on the full January 2024 return series (combined train + test).  GARCH(1,1)-$t$ is trained on the first half and forward evaluated on the test half.  TAR is the ANN lift $\psi \to r$ trained on the train half and forward-evaluated through the Nystr\"om extension on the test half. All values reflect TV distances on a shared $[0,1]$ support. Bootstrap mean $\pm$ standard deviation over $B=200$ resamples of the empirical-test indices.  Bold entries identify the smallest TV in each row.}
\resizebox{\columnwidth}{!}{
	\begin{tabular}{|c|c|c|c|c|}
        \hline
        \textbf{Delay $\tau$} & \textbf{TAR} & \textbf{NIG} & \textbf{Gauss} & \textbf{GARCH(1,1)-$t$} \\
        \hline
        $2$ s   & $0.037 \pm 0.003$& $\mathbf{0.025 \pm 0.005}$ & $0.262 \pm 0.007$ & $0.094 \pm 0.009$ \\
        \hline
        $1$ min & $\mathbf{0.021 \pm 0.002}$ & $0.026 \pm 0.005$ & $0.120 \pm 0.006$ & $0.095 \pm 0.009$ \\
        \hline
        $1$ h   & $\mathbf{0.067 \pm 0.006}$ & $0.112 \pm 0.006$ & $0.119 \pm 0.005$ & $0.207 \pm 0.009$ \\
        \hline
    \end{tabular}}
    \label{tab:L1}
\end{table}

At the mean reverting $\tau$ = 2~s time scale (Fig.\ \ref{TARIII:fig:VOO}(B)(i)),  microstructure effects result in fatter tails well captured by both the NIG model and $TAR_{lift}$ reconstructions, while the Gaussian fit is both too wide and unable to capture fine grain variations. GARCH(1,1)-$t$ falls between the two. At this time scale, the discrete nature of tick size is evident in the empirical returns distribution.  At the intermediate random walk time scale of $\tau$ = 1~min (Fig.\ \ref{TARIII:fig:VOO}(B)(ii)), the $TAR_{lift}$ reconstruction outperforms all parametric models followed by NIG, GARCH(1,1)-$t$, and the Gaussian fit, with the Gaussian fit visibly broader. All models describe the distribution well, but the slight deviation from true random walk behavior exposed by the Hurst exponent of $H \approx 0.48$ means that the higher parameter models, NIG and GARCH(1,1)-$t$, and the nonparametric TAR model offer better fits than the Gaussian. At the longest $\tau$ = 1~h time scale (Fig.\ \ref{TARIII:fig:VOO}(B)(iii)), no parametric model succeeds at recapitulating returns distributions satisfactorily. Only the GARCH(1,1)-$t$ model captures the right-skewness in the returns, and $TAR_{lift}$ offers a 40\% better fit than the best performing parametric model. For the $\tau$ = 2~s and 1~min cases, we observe strong similarity between the test (grey) and train (purple) return distributions, whereas for the $\tau$ = 1~h time scale there is substantial deviation. This indicates that despite the different distributions of the two halves of the data in the trend following regime, the dynamics of the test set returns are adequately represented within the training data to permit accurate $TAR_{lift}$ reconstructions. We note that after sufficiently long time periods have elapsed that the dynamical evolution of the global stock market tracked by VOO has changed enough or moved into a new regime, updating the training data and relearning the manifold $\mathcal{M}^\prime$ would be required. However, over sufficiently short time horizons, a data-driven representation of $\mathcal{M}^\prime$ can be capable of serving as a useful projection for the near-future behaviors of the system as is exploited in empirical dynamical modeling (EDM) \cite{PMID:25733874,Sugiharacritnet}.

This analysis of the January 2024 execution price of shares of the Vanguard S\&P 500 Exchange Traded Fund (VOO) demonstrates the capacity of TAR to recapitulate expected underlying phenomenologies within the data by probing returns over characteristic time scales without access to full-dimensional system observations. TAR provides a single, model-free, geometrically-faithful representation of the return dynamics that admits multi-regime dynamics without imposing any modeling priors, and the learned $\mathcal{M}^\prime$ exposes the heavy-tailed mean reversion, Gaussian random-walk, and skewed trend-following given the relevant $\tau$. One could also employ multi-temporal TAR (Section \ref{TARIII:sec:Villin}) to integrate information across time scales. The learned $\mathcal{M}^\prime$ embedding can also be viewed as a basis in which to perform empirical dynamical modeling \cite{PMID:25733874,Sugiharacritnet} to make short-horizon forecasts, while the projection of new data into $\mathcal{M}^\prime$ using the Nystr\"om extension can be used to identify regime shifts, resolve anomalous excursions into previously upsampled or poorly sampled regions, and identify when retraining of TAR may be warranted.

\section{\label{TARIII:sec:conc} Conclusions and Future Work}

Takens' Theorem provides a powerful and generic theoretical framework to recover low-dimensional manifolds containing the dynamical evolution of a dynamical system from low-dimensional time series in small numbers of system observables \cite{Takens}. By employing modern tools in manifold learning and universal functional approximation, we have previously built upon these mathematical foundations to develop an empirical protocol for data-driven reconstruction of the full-dimensional state of molecular systems from low-dimensional time series \cite{topel2020,topel2023}. In this work, we explored how the delay vector structure impacts reconstruction quality, demonstrated the approach across a variety of dynamical systems. First, we developed a unified, non-parametric workflow termed TAkens Reconstruction (TAR) for the extraction of a geometric intrinsic manifold and reconstruction of system states from low-dimensional time series for arbitrary dynamical systems. Second, we demonstrated in a simple Lotka-Volterra predator-prey model how the time delay can be empirically optimized for accurate system reconstruction. Third, we demonstrated in MD simulations of the protein Villin how the use of multi-temporal delay embeddings presents a numerically efficient means to reconstruct dynamical systems exhibiting multiple characteristic time scales, which, to paraphrase the great Rutherford Aris, allows us ``to get the most out of a time series without really trying'' \cite{aris1976get}. Fourth, we showed in an analysis of historical equity-market return data for which full-dimensional system observations or simulations of the global economy are inaccessible that the learned intrinsic manifold can provide powerful insight into the behaviors of a time series, expose distinct dynamical regimes, recapitulate empirical features of equity-market returns, and predictively reconstruct returns distributions over short time horizons. We have developed a software package entitled TAkens Reconstruction (TAR) to enable users to analyze and reconstruct arbitrary dynamical systems from time series data, which we make available for free and open source download from \url{https://github.com/Ferg-Lab/TAR} and via Zenodo at the persistent DOI \href{https://doi.org/10.5281/zenodo.20115543}{10.5281/zenodo.20115543}.

In future work, we seek to address the interpretability of the learned embeddings to better understand the physical nature of our collective variables by correlating the dimensions of the learned manifold with pools of candidate variables \cite{ma2005automatic} or learning approximate embeddings of the manifold into subspaces spanned by a basis of interpretable variables \cite{kemeth2017equal}. We seek to build upon our study of topological and geometrical nature of time series to understand optimal observable selection to extend our work to non-ergodic systems and explore other dynamical measures of the time series \cite{topel2026}. We also see strong potential to imbue physics-awareness into the analysis through inductive biases or holonomic constraints, where regularizing the data-driven components of TAR through strong priors is anticipated to be particularly useful in data scarce regimes.

\section*{Competing Interests}

\noindent A.L.F.\ is a co-founder and consultant of Evozyne, Inc.\ and a co-author of US Patent Applications 16/887,710 and 17/642,582, US Provisional Patent Applications 62/853,919, 62/900,420, 63/314,898, 63/479,378, 63/479,378, 63/521,617, 63/510,130, 63/669,836, and 63/987,554 and International Patent Applications PCT/US2020/035206, PCT/US2020/050466, PCT/US24/10805, PCT/US24/34369, and PCT/US25/35833.

\section*{Acknowledgements}

\noindent This material is based on the work supported by the National Science Foundation under Grant No.\ DGE-2022023. This work was completed in part with resources provided by the University of Chicago Research Computing Center. We gratefully acknowledge computing time on the University of Chicago high-performance GPU-based cyberinfrastructure supported by the National Science Foundation under Grant No.\ DMR-1828629. This research was supported in part through the computational resources and staff contributions provided for the Quest High-Performance Computing Cluster at Northwestern University, which is jointly supported by the Office of the Provost, the Office for Research, and Northwestern University Information Technology. We thank D.E.~Shaw research for sharing the molecular dynamics simulation trajectories of Villin.

\section*{Data Availability Statement}

\noindent The codes and data used to perform the calculations reported in this paper are made freely available for public access at \url{https://github.com/Ferg-Lab/TAR} and via the persistent Zenodo DOI \href{https://doi.org/10.5281/zenodo.20115543}{10.5281/zenodo.20115543}.

\clearpage
\newpage

\bibliography{ref}

@book{kantz2004nonlinear,
	author = {Holger Kantz and Thomas Schreiber },
	publisher = {Cambridge University Press},
	title = {Nonlinear Time Series Analysis},
	year = {2004}}

@article{martin2024robust,
	author = {Martin, RS and Greve, CM and Huerta, CE and Wong, AS and Koo, JW and Eckhardt, DQ},
	journal = {Chaos: An Interdisciplinary Journal of Nonlinear Science},
	pages = {093110},
	title = {A robust time-delay selection criterion applied to convergent cross mapping},
	volume = {34},
	year = {2024}}

@article{whitney1936differentiable,
	author = {Whitney, Hassler},
	journal = {Annals of Mathematics},
	number = {3},
	pages = {645--680},
	publisher = {JSTOR},
	title = {Differentiable manifolds},
	volume = {37},
	year = {1936}}

@article{das2006low,
	author = {Das, Payel and Moll, Mark and Stamati, Hernan and Kavraki, Lydia E and Clementi, Cecilia},
	journal = {Proceedings of the National Academy of Sciences of the United States of America},
	number = {26},
	pages = {9885--9890},
	publisher = {National Academy of Sciences},
	title = {Low-dimensional, free-energy landscapes of protein-folding reactions by nonlinear dimensionality reduction},
	volume = {103},
	year = {2006}}

@article{zhuravlev2009deconstructing,
	author = {Zhuravlev, Pavel I and Materese, Christopher Kroboth and Papoian, Garegin A},
	journal = {The Journal of Physical Chemistry B},
	number = {26},
	pages = {8800--8812},
	publisher = {ACS Publications},
	title = {Deconstructing the native state: energy landscapes, function, and dynamics of globular proteins},
	volume = {113},
	year = {2009}}

@article{hegger2007complex,
	author = {Hegger, Rainer and Altis, Alexandros and Nguyen, Phuong H and Stock, Gerhard},
	journal = {Physical Review Letters},
	number = {2},
	pages = {028102},
	publisher = {APS},
	title = {How complex is the dynamics of peptide folding?},
	volume = {98},
	year = {2007}}

@article{amadei1993essential,
	author = {Amadei, Andrea and Linssen, Antonius BM and Berendsen, Herman JC},
	journal = {Proteins: Structure, Function, and Bioinformatics},
	number = {4},
	pages = {412--425},
	publisher = {Wiley Online Library},
	title = {Essential dynamics of proteins},
	volume = {17},
	year = {1993}}

@article{garcia1992large,
	author = {Garc{\'\i}a, Angel E},
	journal = {Physical Review Letters},
	number = {17},
	pages = {2696},
	publisher = {APS},
	title = {Large-amplitude nonlinear motions in proteins},
	volume = {68},
	year = {1992}}

@article{topel2026,
	author = {Topel, Maximilian},
	journal = {arXiv preprint arXiv:2604.27412},
	title = {{Kolmogorov-Sinai} entropies identify optimal observables for prediction and dynamics reconstruction in chaotic systems},
	year = {2026}}

@article{kemeth2017equal,
	author = {Kemeth, F. P. and Haugland, S. W. and Dietrich, F. and Bertalan, T. and Li, Q. and Bollt, E. M. and Talmon, R. and Krischer, K. and Kevrekidis, Ioannis G.},
	journal = {arXiv preprint arXiv.1708.05406},
	title = {An equal space for complex data with unknown internal order: Observability, gauge invariance and manifold learning},
	year = {2017}}

@article{aris1976get,
	author = {Aris, Rutherford},
	journal = {Chemical Engineering Education},
	number = {3},
	pages = {114--124},
	title = {How to get the most out of an equation without really trying},
	volume = {10},
	year = {1976}}

@incollection{junghare2023markov,
	author = {Junghare, Vivek and Bhattacharya, Sourya and Ansari, Khalid and Hazra, Saugata},
	booktitle = {Protein Folding Dynamics and Stability: Experimental and Computational Methods},
	doi = {10.1007/978-981-99-2079-2_8},
	pages = {147--164},
	publisher = {Springer Nature Singapore},
	title = {Markov State Models of Molecular Simulations to Study Protein Folding and Dynamics},
	year = {2023}}

@article{sidky2020molecular,
	author = {Sidky, Hythem and Chen, Wei and Ferguson, Andrew L},
	doi = {10.1039/D0SC03635H},
	journal = {Chemical Science},
	number = {35},
	pages = {9459--9467},
	publisher = {Royal Society of Chemistry},
	title = {Molecular latent space simulators},
	volume = {11},
	year = {2020}}

@article{jones2025flowback,
	author = {Jones, Michael S. and Khanna, Smayan and Ferguson, Andrew L.},
	doi = {10.1021/acs.jcim.4c02046},
	journal = {Journal of Chemical Information and Modeling},
	number = {2},
	pages = {672--692},
	title = {{FlowBack}: A Generalized Flow-Matching Approach for Biomolecular Backmapping},
	volume = {65},
	year = {2025}}

@article{jones2023diamondback,
	author = {Jones, Michael S. and Shmilovich, Kirill and Ferguson, Andrew L.},
	doi = {10.1021/acs.jctc.3c00840},
	journal = {Journal of Chemical Theory and Computation},
	number = {21},
	pages = {7908--7923},
	publisher = {ACS Publications},
	title = {{DiAMoNDBack}: Diffusion-denoising Autoregressive Model for Non-Deterministic Backmapping of {C$\alpha$} Protein Traces},
	volume = {19},
	year = {2023}}

@article{kevrekidis2009equation,
	author = {Kevrekidis, Ioannis G and Samaey, Giovanni},
	doi = {10.1146/annurev.physchem.59.032607.093610},
	journal = {Annual Review of Physical Chemistry},
	pages = {321--344},
	publisher = {Annual Reviews},
	title = {Equation-free multiscale computation: Algorithms and applications},
	volume = {60},
	year = {2009}}

@article{wang2018study,
	author = {Wang, Jiang and Ferguson, Andrew L.},
	doi = {10.1021/acs.macromol.7b01684},
	journal = {Macromolecules},
	number = {2},
	pages = {598--616},
	publisher = {ACS Publications},
	title = {A Study of the Morphology, Dynamics, and Folding Pathways of Ring Polymers with Supramolecular Topological Constraints Using Molecular Simulation and Nonlinear Manifold Learning},
	volume = {51},
	year = {2018}}

@article{Schulten:96:JMG,
	author = {Humphrey, William and Dalke, Andrew and Schulten, Klaus},
	doi = {10.1016/0263-7855(96)00018-5},
	journal = {Journal of Molecular Graphics},
	number = {1},
	pages = {33--38},
	publisher = {Elsevier},
	title = {{VMD}: Visual molecular dynamics},
	volume = {14},
	year = {1996}}

@article{Ferg16,
	author = {Wang, J. and Ferguson, A. L.},
	doi = {10.1103/PhysRevE.93.032412},
	journal = {Phys. Rev. E},
	pages = {032412},
	title = {Nonlinear reconstruction of single-molecule free energy surfaces from univariate time series},
	volume = {93},
	year = {2016}}

@article{Ferg18,
	author = {Wang, J. and Ferguson, A. L.},
	doi = {10.1021/acs.jpcb.8b08800},
	journal = {J. Phys. Chem. B},
	number = {50},
	pages = {11931--11952},
	title = {Recovery of protein folding funnels from single-molecule time series by delay embeddings and manifold learning},
	volume = {122},
	year = {2018}}

@inbook{EQfree,
	address = {Dordrecht},
	author = {Kevrekidis, Ioannis G. and Gear, C. William and Hummer, Gerhard},
	booktitle = {Handbook of Materials Modeling: Methods},
	doi = {10.1007/978-1-4020-3286-8_72},
	pages = {1453--1475},
	publisher = {Springer Netherlands},
	title = {Equation-free Modeling For Complex Systems},
	year = {2005}}

@article{Sugiharacritnet,
	author = {Scheffer, Marten and Bascompte, Jordi and Brock, William A. and Brovkin, Victor and Carpenter, Stephen R. and Dakos, Vasilis and Held, Hermann and van Nes, Egbert H. and Rietkerk, Max and Sugihara, George},
	doi = {10.1038/nature08227},
	journal = {Nature},
	number = {7260},
	pages = {53--59},
	title = {Early-warning signals for critical transitions},
	volume = {461},
	year = {2009}}

@article{hurst1956p1,
	author = {Hurst, H. E.},
	doi = {10.1080/02626665609493644},
	journal = {International Association of Scientific Hydrology. Bulletin},
	number = {3},
	pages = {13--27},
	title = {The problem of long-term storage in reservoirs},
	volume = {1},
	year = {1956}}

@article{hurst1957p2,
	author = {Hurst, H. E.},
	doi = {10.1038/180494a0},
	journal = {Nature},
	pages = {494},
	title = {A suggested statistical model of some time series which occur in nature},
	volume = {180},
	year = {1957}}

@article{weron2002hurst,
	author = {Weron, R.},
	doi = {10.1016/S0378-4371(02)00961-5},
	journal = {Physica A: Statistical Mechanics and its Applications},
	number = {1},
	pages = {285--299},
	title = {Estimating long-range dependence: Finite sample properties and confidence intervals},
	volume = {312},
	year = {2002}}

@book{econometricsfinmark,
	address = {Princeton, NJ},
	author = {Campbell, John Y. and Lo, Andrew W. and MacKinlay, A. Craig},
	isbn = {9780691043012},
	publisher = {Princeton University Press},
	title = {The Econometrics of Financial Markets},
	year = {1997}}

@book{Bouchaud_TQP,
	author = {Bouchaud, Jean-Philippe and Bonart, Julius and Donier, Jonathan and Gould, Martin},
	doi = {10.1017/9781316659335},
	publisher = {Cambridge University Press},
	title = {Trades, Quotes and Prices: Financial Markets Under the Microscope},
	year = {2018}}

@book{OharaMicro,
	author = {O'Hara, Maureen},
	publisher = {Blackwell Publishers},
	title = {Market Microstructure Theory},
	year = {1995}}

@article{RVHansenLunde2006,
	author = {Hansen, Peter Reinhard and Lunde, Asger},
	doi = {10.1198/073500106000000071},
	journal = {Journal of Business \& Economic Statistics},
	number = {2},
	pages = {127--161},
	title = {Realized Variance and Market Microstructure Noise},
	volume = {24},
	year = {2006}}

@article{MADHAVANmicrostructure,
	author = {Ananth Madhavan},
	doi = {10.1016/S1386-4181(00)00007-0},
	journal = {Journal of Financial Markets},
	number = {3},
	pages = {205-258},
	title = {Market microstructure: A survey},
	volume = {3},
	year = {2000}}

@misc{WRDS,
	author = {{Wharton Research Data Services}},
	note = {Dataset retrieved from Wharton Research Data Services (https://wrds-www.wharton.upenn.edu/); all executed sales transactions during trading hours in January 2024},
	title = {Executed Trades Data for Vanguard {S\&P} 500 {ETF} ({VOO})},
	year = {2024}}

@incollection{kleinberg2010networks,
	author = {Easley, David and Kleinberg, Jon},
	booktitle = {Networks, Crowds, and Markets: Reasoning about a Highly Connected World},
	chapter = {16},
	doi = {10.1017/CBO9780511761942},
	pages = {483--508},
	publisher = {Cambridge University Press},
	title = {Information Cascades},
	year = {2010}}

@article{networkcascade,
	author = {Brummitt, Charles D. and Lee, Kyu-Min and Goh, K.-I.},
	doi = {10.1103/PhysRevE.85.045102},
	journal = {Phys. Rev. E},
	number = {4},
	pages = {045102},
	publisher = {American Physical Society},
	title = {Multiplexity-facilitated cascades in networks},
	volume = {85},
	year = {2012}}

@book{Haken1983,
	author = {Haken, Hermann},
	doi = {10.1007/978-3-642-88338-5},
	publisher = {Springer Berlin Heidelberg},
	title = {Synergetics: An Introduction},
	year = {1983}}

@article{proteinTS,
	author = {Henzler-Wildman, Katherine and Kern, Dorothee},
	doi = {10.1038/nature06522},
	journal = {Nature},
	number = {7172},
	pages = {964--972},
	title = {Dynamic personalities of proteins},
	volume = {450},
	year = {2007}}

@article{deshaw,
	author = {Lindorff-Larsen, K. and Piana, S. and Dror, R. O. and Shaw, D. E.},
	doi = {10.1126/science.1208351},
	journal = {Science},
	number = {6055},
	pages = {517--520},
	title = {How Fast-Folding Proteins Fold},
	volume = {334},
	year = {2011}}

@article{stark2003delay,
	author = {Stark, J. and Broomhead, David S. and Davies, M. E. and Huke, J.},
	doi = {10.1007/s00332-003-0534-4},
	journal = {Journal of Nonlinear Science},
	number = {6},
	pages = {519--577},
	publisher = {Springer},
	title = {Delay embeddings for forced systems. {II}. {S}tochastic forcing},
	volume = {13},
	year = {2003}}

@article{stark1999delay,
	author = {Stark, Jaroslav},
	doi = {10.1007/s003329900072},
	journal = {Journal of Nonlinear Science},
	number = {3},
	pages = {255--332},
	publisher = {Springer},
	title = {Delay embeddings for forced systems. {I}. {D}eterministic forcing},
	volume = {9},
	year = {1999}}

@incollection{nadler2006advances,
	author = {Nadler, B. and Lafon, S. and Coifman, R. R. and Kevrekidis, I. G.},
	booktitle = {Advances in Neural Information Processing Systems 18: Proceedings of the 2005 Conference (Neural Information Processing)},
	pages = {955-962},
	publisher = {The MIT Press},
	title = {Diffusion Maps, Spectral Clustering and Eigenfunctions of {F}okker-{P}lanck Operators},
	year = {2006}}

@inproceedings{nystrom,
	author = {Williams, Christopher and Seeger, Matthias},
	booktitle = {Advances in Neural Information Processing Systems},
	pages = {682--688},
	publisher = {MIT Press},
	title = {Using the {Nystr\"o}m Method to Speed Up Kernel Machines},
	volume = {13},
	year = {2000}}

@incollection{NLPCA,
	author = {Scholz, Matthias and Fraunholz, Martin and Selbig, Joachim},
	booktitle = {Principal Manifolds for Data Visualization and Dimension Reduction},
	doi = {10.1007/978-3-540-73750-6_2},
	pages = {44-67},
	publisher = {Springer Berlin Heidelberg},
	series = {Lecture Notes in Computational Science and Engineering},
	title = {Nonlinear Principal Component Analysis: {N}eural Network Models and Applications},
	volume = {58},
	year = {2008}}

@article{RWFama,
	author = {Eugene F. Fama},
	doi = {10.2469/faj.v51.n1.1861},
	journal = {Financial Analysts Journal},
	number = {1},
	pages = {75--80},
	publisher = {Routledge},
	title = {Random Walks in Stock Market Prices},
	volume = {51},
	year = {1995}}

@article{sauer1991embedology,
	author = {Sauer, Tim and Yorke, James A and Casdagli, Martin},
	doi = {10.1007/BF01053745},
	journal = {Journal of Statistical Physics},
	number = {3-4},
	pages = {579--616},
	publisher = {Springer},
	title = {Embedology},
	volume = {65},
	year = {1991}}

@article{RuelleEckmann,
	author = {Eckmann, J. -P. and Ruelle, D.},
	doi = {10.1103/RevModPhys.57.617},
	journal = {Reviews of Modern Physics},
	number = {3},
	pages = {617--656},
	publisher = {American Physical Society},
	title = {Ergodic theory of chaos and strange attractors},
	volume = {57},
	year = {1985}}

@article{ma2005automatic,
	author = {Ma, Ao and Dinner, Aaron R},
	doi = {10.1021/jp045546c},
	journal = {Journal of Physical Chemistry B},
	number = {14},
	pages = {6769--6779},
	publisher = {ACS Publications},
	title = {Automatic method for identifying reaction coordinates in complex systems},
	volume = {109},
	year = {2005}}

@inproceedings{salvador2004determining,
	author = {Salvador, Stan and Chan, Philip},
	booktitle = {16th IEEE International Conference on Tools with Artificial Intelligence (ICTAI), 2004},
	doi = {10.1109/ICTAI.2004.50},
	pages = {576-584},
	title = {Determining the number of clusters/segments in hierarchical clustering/segmentation algorithms},
	year = {2004}}

@article{coifman2006diffusion,
	author = {Coifman, Ronald R. and Lafon, St{\'e}phane},
	doi = {10.1016/j.acha.2006.04.006},
	journal = {Applied and Computational Harmonic Analysis},
	number = {1},
	pages = {5--30},
	publisher = {Elsevier},
	title = {Diffusion maps},
	volume = {21},
	year = {2006}}

@article{coifman2008diffusion,
	author = {Coifman, Ronald R and Kevrekidis, Ioannis G and Lafon, St{\'e}phane and Maggioni, Mauro and Nadler, Boaz},
	doi = {10.1137/070696325},
	journal = {Multiscale Modeling \& Simulation},
	number = {2},
	pages = {842--864},
	publisher = {SIAM},
	title = {Diffusion maps, reduction coordinates, and low dimensional representation of stochastic systems},
	volume = {7},
	year = {2008}}

@article{lpbeltrami,
	author = {Nadler, Boaz and Lafon, St{\'e}phane and Coifman, Ronald R and Kevrekidis, Ioannis G},
	doi = {10.1016/j.acha.2005.07.004},
	journal = {Applied and Computational Harmonic Analysis},
	number = {1},
	pages = {113-127},
	title = {Diffusion maps, spectral clustering and reaction coordinates of dynamical systems},
	volume = {21},
	year = {2006}}

@article{ferguson2011cpl,
	author = {Ferguson, Andrew L and Panagiotopoulos, Athanassios Z and Kevrekidis, Ioannis G and Debenedetti, Pablo G},
	journal = {Chemical Physics Letters},
	pages = {1--11},
	title = {Nonlinear dimensionality reduction in molecular simulation: The diffusion map approach},
	volume = {509},
	year = {2011}}

@article{topel2023,
	author = {Topel, Maximilian and Ejaz, Ayesha and Squires, Allison and Ferguson, Andrew L.},
	doi = {10.1021/acs.jctc.2c00920},
	journal = {Journal of Chemical Theory and Computation},
	number = {14},
	pages = {4654-4667},
	title = {Learned Reconstruction of Protein Folding Trajectories from Noisy Single-Molecule Time Series},
	volume = {19},
	year = {2023}}

@article{topel2020,
	author = {Topel, Maximilian and Ferguson, Andrew L.},
	doi = {10.1063/5.0024732},
	journal = {The Journal of Chemical Physics},
	number = {19},
	pages = {194102},
	title = {{Reconstruction of protein structures from single-molecule time series}},
	volume = {153},
	year = {2020}}

@article{ferguson2010systematic,
	author = {Ferguson, Andrew L and Panagiotopoulos, Athanassios Z and Debenedetti, Pablo G and Kevrekidis, Ioannis G},
	doi = {10.1073/pnas.1003293107},
	journal = {Proceedings of the National Academy of Sciences of the United States of America},
	number = {31},
	pages = {13597--13602},
	publisher = {National Acad Sciences},
	title = {Systematic determination of order parameters for chain dynamics using diffusion maps},
	volume = {107},
	year = {2010}}

@article{LLE,
	author = {Sam T. Roweis and Lawrence K. Saul},
	doi = {10.1126/science.290.5500.2323},
	journal = {Science},
	number = {5500},
	pages = {2323-2326},
	title = {Nonlinear Dimensionality Reduction by Locally Linear Embedding},
	volume = {290},
	year = {2000}}

@article{ISOMAP,
	author = {Joshua B. Tenenbaum and Vin de Silva and John C. Langford},
	doi = {10.1126/science.290.5500.2319},
	journal = {Science},
	number = {5500},
	pages = {2319-2323},
	title = {A Global Geometric Framework for Nonlinear Dimensionality Reduction},
	volume = {290},
	year = {2000}}

@article{LV,
	author = {Peter J. Wangersky},
	doi = {10.1146/annurev.es.09.110178.001201},
	journal = {Annual Review of Ecology and Systematics},
	pages = {189--218},
	publisher = {Annual Reviews},
	title = {{Lotka-Volterra} Population Models},
	volume = {9},
	year = {1978}}

@article{Roy2008rr,
	author = {Roy, Rahul and Hohng, Sungchul and Ha, Taekjip},
	doi = {10.1038/nmeth.1208},
	journal = {Nature Methods},
	number = {6},
	pages = {507--516},
	publisher = {Nature Publishing Group},
	title = {A practical guide to single-molecule {FRET}},
	volume = {5},
	year = {2008}}

@book{difftop,
	author = {Hirsch, Morris W.},
	doi = {10.1007/978-1-4684-9449-5},
	number = {33},
	publisher = {Springer-Verlag},
	series = {Graduate Texts in Mathematics},
	title = {Differential Topology},
	year = {1994}}

@article{Broomhead1986,
	author = {Broomhead, D S and King, Gregory P},
	doi = {10.1016/0167-2789(86)90031-X},
	journal = {Physica D: Nonlinear Phenomena},
	number = {2--3},
	pages = {217--236},
	title = {{Extracting qualitative dynamics from experimental data}},
	volume = {20},
	year = {1986}}

@article{Cao1998,
	author = {Cao, Liangyue and Mees, Alistair and Judd, Kevin},
	doi = {10.1016/S0167-2789(98)00151-1},
	journal = {Physica D: Nonlinear Phenomena},
	number = {1--2},
	pages = {75--88},
	title = {Dynamics from multivariate time series},
	volume = {121},
	year = {1998}}

@article{probtak,
	author = {Krzysztof Bara{\'n}ski and Yonatan Gutman and Adam {\'S}piewak},
	doi = {10.1088/1361-6544/ab8fb8},
	journal = {Nonlinearity},
	number = {9},
	pages = {4940},
	publisher = {IOP Publishing},
	title = {A probabilistic Takens theorem},
	volume = {33},
	year = {2020}}

@article{Coifman24052005,
	author = {Coifman, R R and Lafon, S and Lee, A B and Maggioni, M and Nadler, B and Warner, F and Zucker, S W},
	doi = {10.1073/pnas.0500334102},
	journal = {Proceedings of the National Academy of Sciences of the United States of America},
	number = {21},
	pages = {7426--7431},
	title = {Geometric diffusions as a tool for harmonic analysis and structure definition of data: Diffusion maps},
	volume = {102},
	year = {2005}}

@article{packard1980,
	author = {Packard, N. H. and Crutchfield, J. P. and Farmer, J. D. and Shaw, R. S.},
	doi = {10.1103/PhysRevLett.45.712},
	journal = {Physical Review Letters},
	number = {9},
	pages = {712--716},
	title = {Geometry from a Time Series},
	volume = {45},
	year = {1980}}

@incollection{Takens,
	address = {Berlin, Heidelberg},
	author = {Takens, Floris},
	booktitle = {Dynamical Systems and Turbulence, Warwick 1980},
	doi = {10.1007/BFb0091924},
	editor = {Rand, D. A. and Young, L.-S.},
	pages = {366--381},
	publisher = {Springer Berlin Heidelberg},
	series = {Lecture Notes in Mathematics},
	title = {Detecting strange attractors in turbulence},
	volume = {898},
	year = {1981}}

@article{Wu1994,
	author = {Wu, P G and Brand, L},
	doi = {10.1006/abio.1994.1134},
	journal = {Analytical Biochemistry},
	pages = {1--13},
	title = {Resonance Energy Transfer: Methods and Applications},
	volume = {218},
	year = {1994}}

@article{Cao97,
	author = {Liangyue Cao},
	doi = {10.1016/S0167-2789(97)00118-8},
	journal = {Physica D: Nonlinear Phenomena},
	number = {1--2},
	pages = {43--50},
	title = {Practical method for determining the minimum embedding dimension of a scalar time series},
	volume = {110},
	year = {1997}}

@article{MI,
	author = {Fraser, Andrew M. and Swinney, Harry L.},
	doi = {10.1103/PhysRevA.33.1134},
	journal = {Physical Review A},
	number = {2},
	pages = {1134--1140},
	publisher = {American Physical Society},
	title = {Independent coordinates for strange attractors from mutual information},
	volume = {33},
	year = {1986}}

@article{PMID:25733874,
	author = {Ye, Hao and Beamish, Richard J and Glaser, Sarah M and Grant, Sue C H and Hsieh, Chih-Hao and Richards, Laura J and Schnute, Jon T and Sugihara, George},
	doi = {10.1073/pnas.1417063112},
	journal = {Proceedings of the National Academy of Sciences of the United States of America},
	number = {13},
	pages = {E1569---76},
	title = {Equation-free mechanistic ecosystem forecasting using empirical dynamic modeling},
	volume = {112},
	year = {2015}}

@article{Hollingsworth2018,
	author = {Hollingsworth, Scott A. and Dror, Ron O.},
	doi = {10.1016/j.neuron.2018.08.011},
	journal = {Neuron},
	number = {6},
	pages = {1129-1143},
	publisher = {Elsevier},
	title = {Molecular Dynamics Simulation for All},
	volume = {99},
	year = {2018}}

@book{Hassoun1996,
	address = {Cambridge, MA},
	author = {Hassoun, Mohamad H.},
	isbn = {9780262082396},
	publisher = {MIT Press},
	title = {Fundamentals of Artificial Neural Networks},
	year = {1995}}

@article{keverkidis2017,
	author = {Or Yair and Ronen Talmon and Ronald R. Coifman and Ioannis G. Kevrekidis},
	doi = {10.1073/pnas.1620045114},
	journal = {Proceedings of the National Academy of Sciences of the United States of America},
	number = {38},
	pages = {E7865-E7874},
	title = {Reconstruction of normal forms by learning informed observation geometries from data},
	volume = {114},
	year = {2017}}

@book{strogatz,
	author = {Strogatz, Steven H.},
	publisher = {CRC Press},
	title = {Nonlinear Dynamics and Chaos: With Applications to Physics, Biology, Chemistry, and Engineering},
	year = {2024}}

@book{Brin_Stuck_2002,
	author = {Brin, Michael and Stuck, Garrett},
	doi = {10.1017/CBO9780511755316},
	publisher = {Cambridge University Press},
	title = {Introduction to Dynamical Systems},
	year = {2002}}

@article{Suddhasattwa2023,
	author = {Berry, Tyrus and Das, Suddhasattwa},
	doi = {10.1137/22M1516865},
	journal = {SIAM Journal on Applied Dynamical Systems},
	number = {3},
	pages = {2082-2122},
	title = {Learning Theory for Dynamical Systems},
	volume = {22},
	year = {2023}}

@article{Sugihara2011PLOS,
	author = {Deyle, Ethan R. AND Sugihara, George},
	doi = {10.1371/journal.pone.0018295},
	journal = {PLOS ONE},
	number = {3},
	pages = {1-8},
	publisher = {Public Library of Science},
	title = {Generalized Theorems for Nonlinear State Space Reconstruction},
	volume = {6},
	year = {2011}}

@article{Solijacic2022,
	author = {Lu, Peter Y. and Ari{\~n}o Bernad, Joan and Solja{\v c}i{\'c}, Marin},
	doi = {10.1038/s42005-022-00987-z},
	journal = {Communications Physics},
	number = {1},
	pages = {206},
	title = {Discovering sparse interpretable dynamics from partial observations},
	volume = {5},
	year = {2022}}

@article{Butcher_1964,
	author = {Butcher, J. C.},
	doi = {10.1017/S1446788700023387},
	journal = {Journal of the Australian Mathematical Society},
	number = {2},
	pages = {179--194},
	title = {On {Runge-Kutta} processes of high order},
	volume = {4},
	year = {1964}}

@article{landmarkdiffusion,
	author = {Andrew W. Long and Andrew L. Ferguson},
	doi = {10.1016/j.acha.2017.08.004},
	journal = {Applied and Computational Harmonic Analysis},
	number = {1},
	pages = {190-211},
	title = {Landmark diffusion maps ({L-dMaps}): Accelerated manifold learning out-of-sample extension},
	volume = {47},
	year = {2019}}

@article{baker2020covid,
	author = {Baker, Scott R. and Bloom, Nicholas and Davis, Steven J. and Kost, Kyle and Sammon, Marco and Viratyosin, Tasaneeya},
	doi = {10.1093/rapstu/raaa008},
	journal = {The Review of Asset Pricing Studies},
	number = {4},
	pages = {742--758},
	title = {The unprecedented stock market reaction to {COVID-19}},
	volume = {10},
	year = {2020}}

@article{taqqu1995estimators,
	author = {Taqqu, Murad S. and Teverovsky, Vadim and Willinger, Walter},
	doi = {10.1142/S0218348X95000692},
	journal = {Fractals},
	number = {4},
	pages = {785--798},
	title = {Estimators for long-range dependence: An empirical study},
	volume = {3},
	year = {1995}}

@article{barndorffnielsen1997nig,
	author = {Barndorff-Nielsen, Ole E.},
	doi = {10.1111/1467-9469.00045},
	journal = {Scandinavian Journal of Statistics},
	number = {1},
	pages = {1--13},
	title = {Normal Inverse {G}aussian Distributions and Stochastic Volatility Modelling},
	volume = {24},
	year = {1997}}

@article{clark1973time,
	author = {Clark, Peter K.},
	doi = {10.2307/1913889},
	journal = {Econometrica},
	number = {1},
	pages = {135--155},
	title = {A Subordinated Stochastic Process Model with Finite Variance for Speculative Prices},
	volume = {41},
	year = {1973}}

@article{roll1984spread,
	author = {Roll, Richard},
	doi = {10.1111/j.1540-6261.1984.tb03897.x},
	journal = {The Journal of Finance},
	number = {4},
	pages = {1127--1139},
	title = {A Simple Implicit Measure of the Effective Bid--Ask Spread in an Efficient Market},
	volume = {39},
	year = {1984}}

@article{bollerslev1986garch,
	author = {Bollerslev, Tim},
	doi = {10.1016/0304-4076(86)90063-1},
	journal = {Journal of Econometrics},
	number = {3},
	pages = {307--327},
	title = {Generalized Autoregressive Conditional Heteroskedasticity},
	volume = {31},
	year = {1986}}

@article{bollerslev1987tgarch,
	author = {Bollerslev, Tim},
	doi = {10.2307/1925546},
	journal = {The Review of Economics and Statistics},
	number = {3},
	pages = {542--547},
	title = {A Conditionally Heteroskedastic Time Series Model for Speculative Prices and Rates of Return},
	volume = {69},
	year = {1987}}

@article{mandelbrot1963variation,
	author = {Mandelbrot, Benoit},
	doi = {10.1086/294632},
	journal = {The Journal of Business},
	number = {4},
	pages = {394--419},
	title = {The Variation of Certain Speculative Prices},
	volume = {36},
	year = {1963}}

@article{fama1965behavior,
	author = {Fama, Eugene F.},
	doi = {10.1086/294743},
	journal = {The Journal of Business},
	number = {1},
	pages = {34--105},
	title = {The Behavior of Stock-Market Prices},
	volume = {38},
	year = {1965}}

@article{engle1982arch,
	author = {Engle, Robert F.},
	doi = {10.2307/1912773},
	journal = {Econometrica},
	number = {4},
	pages = {987--1007},
	title = {Autoregressive Conditional Heteroscedasticity with Estimates of the Variance of {United Kingdom} Inflation},
	volume = {50},
	year = {1982}}

@article{hotelling1936,
	author = {Hotelling, Harold},
	journal = {Biometrika},
	number = {3/4},
	pages = {321--377},
	title = {Relations between two sets of variates},
	volume = {28},
	year = {1936}}

\end{document}